\documentclass[twocolumn,reprint,noshowpacs]{revtex4-1}
\usepackage{graphicx}
\usepackage{amsmath}
\usepackage{amssymb}
\usepackage{epsfig}

 \usepackage[
         pdftitle={Adsorption of finite semiflexible polymers and 
 their loop and tail distributions},
          pdfauthor={Tobias Kampmann, Jan Kierfeld}
          ]{hyperref}

\usepackage[utf8]{inputenc}
\usepackage{graphicx}
\usepackage{placeins}
\usepackage{lastpage}
\usepackage{mathtools}
\usepackage{amsmath,amssymb}
\usepackage[format=plain,justification=raggedright,singlelinecheck=false,
font=small,labelfont=bf,labelsep=space]{caption} 
\usepackage{fancyhdr}
\usepackage{upgreek}
 \let \Beta B

 \usepackage{color,colordvi}
 \definecolor{Red}{rgb}{0.9,0.0,0.1}

\let \phi \varphi

\let \pi \uppi
\usepackage{xcolor}
\newcommand{\br}{{\bf r}}

\newcommand{\A}{\textbf{\textsf{A}}\ }
\newcommand{\B}{\textbf{\textsf{B}}\ }

\begin{document}

 \title{Adsorption of finite semiflexible polymers and 
 their loop and tail distributions}

 \author{Tobias A. Kampmann}
 \email{tobias.kampmann@udo.edu}
 \author{Jan Kierfeld}
 \email{jan.kierfeld@tu-dortmund.de}
 \affiliation{Physics Department, TU Dortmund University, 
 44221 Dortmund, Germany}

 \date{\today}

 \begin{abstract}
We discuss the adsorption of semiflexible polymers to a planar
 attractive wall 
and focus on the questions of the adsorption threshold for 
polymers of {\it finite} length and their loop and tail distributions 
 using both Monte-Carlo simulations and 
analytical arguments. 
For the adsorption threshold, we find three regimes:
(i) a  flexible or 
Gaussian regime if the persistence length is smaller than the adsorption 
potential range, (ii) a semiflexible regime if the persistence length is 
larger than the potential range, and (iii) for finite polymers, 
a novel crossover to a  rigid rod 
regime if the  deflection length exceeds the contour length. 
 In the flexible and semiflexible regime, finite size corrections 
arise because  the correlation length exceeds the 
contour length. 
In the rigid rod regime, however,  it is essential how 
 the global orientational  or translational 
degrees of freedom are restricted by grafting or confinement.
We discuss finite size corrections 
 for polymers grafted to the adsorbing surface and 
for polymers confined by a second (parallel) hard wall. 
Based on these results we obtain a method to analyze 
adsorption data for finite semiflexible polymers
such as filamentous actin.
For the loop and tail distributions, we find 
 power laws with an exponential decay on length
scales exceeding the correlation length.
We derive and confirm the loop and tail power law exponents
for flexible and semiflexible polymers. 
This allows us to explain that, close to the transition, 
 semiflexible polymers have significantly smaller loops and 
both flexible and semiflexible polymers desorb by expanding their 
tail length. 
The  tail distribution 
allows us to extract the free energy per length  of adsorption 
for 
actin filaments from experimental data 
[D. Welch {\it et al.},  Soft Matter {\bf 11}, 7507 (2015)].
 \end{abstract}

 \pacs{???}

 \maketitle 

\section{Introduction}
\footnotetext{\textit{Physics Department, TU Dortmund University, 
44221 Dortmund, Germany; E-mail: jan.kierfeld@tu-dortmund.de}}

For semiflexible polymers 
their intrinsic 
bending energy associated with their bending rigidity $\kappa$
 is relevant for 
shape fluctuations. The competition 
between thermal and bending energy determines the 
persistence length $L_p\sim \kappa/k_BT$ of the polymer, which 
is the decay length of orientational correlations along a free
polymer. For semiflexible polymers the persistence length is 
large and comparable to other length scales in the problem.
 Examples of semiflexible polymers are 
 stiff synthetic polymers such as polyelectrolytes
\cite{Skolnick1977,Netz1999} and  many stiff biopolymers 
such as DNA, filamentous (F-)actin, or microtubules.
The persistence length of  F-actin 
is in the $10\,{\rm \mu m}$-range  \cite{ott1993}
and ranges up  to the mm-range for
 microtubules \cite{venier1994}.
The bending rigidity also modifies the
 adsorption behavior of semiflexible polymers.  
In a recent experiment \cite{Welch2015}, the conformations 
of single, finite actin filaments adsorbed onto a planar wall by a 
depletion interaction have been analyzed
and  compared   with Monte-Carlo simulations based on a
phenomenological treatment of finite size effects.
In this paper we want to go beyond the  analysis 
given in Ref.\ \citenum{Welch2015} and systematically 
derive a procedure to analyze
finite size effects for semiflexible polymer adsorption.

The  adsorption of single flexible 
 polymer chains has been extensively studied 
theoretically (see, for example, Refs.\
 \citenum{degennes,eisenriegler,Netz2003}).
Theoretical studies on the 
  adsorption of semiflexible polymers with intrinsic bending stiffness 
are less but still numerous
\cite{Birshtein1979,Maggs1989,Gompper1989,Gompper1990,Khokhlov1993,kramarenko96,VanderLinden1996,Kuznetsov1997,Netz1999,bundschuh00,sintes01,Stepanow2001,Semenov2002,kierfeld03,Benetatos2003,Kierfeld2006,owczarek2009,deng10,Kampmann2013,Hsu2013,Welch2015}
and fall into two main classes, which are  
studies of lattice polymers
\cite{Birshtein1979,Khokhlov1993,VanderLinden1996,owczarek2009,Hsu2013} or 
of off-lattice continuous polymers
\cite{Maggs1989,Gompper1989,Gompper1990,kramarenko96,Kuznetsov1997,Netz1999,bundschuh00,sintes01,Stepanow2001,Semenov2002,kierfeld03,Kierfeld2006,deng10,Kampmann2013,Welch2015};
in Ref.\ \citenum{Benetatos2003}, only the binding potential was realized in 
a discrete manner by explicitly discrete pinning sites. 
We will conduct off-lattice simulations and exact off-lattice 
calculations in the rigid rod limit. 
In order  to quantify loop and tail distributions we 
use critical exponent relations based 
on  necklace  model  \cite{Fisher1984} and transfer matrix approaches
for off-lattice polymers.
Closely related to adsorption is 
 the conformational statistics of semiflexible polymers confined to 
the half-plane \cite{burkhardt} and  slits or channels
\cite{Odijk1983,harnau1999,Bicout2001,Kraikivski2005a,Levi2007,Koster2007,Koster2008},
from which we will also use concepts such as the deflection length 
and exact results on critical exponents.

From a theoretical point of view, the adsorption of semiflexible 
 polymers is challenging because it involves  several competing 
 length scales. For a freely fluctuating 
 semiflexible polymer, the 
 persistence length $L_p$ and its contour length $L$ are the relevant length 
 scales. For $L\lesssim L_p$, thermal fluctuations are 
dominated by bending energy. This is the regime we will mostly 
focus on in this work and which is relevant for 
actin filaments, where typically  both $L$  and  $L_p$ 
 are  in the range of  $10-20\,{\rm \mu m}$.
For $L\lesssim L_p$ the bending energy will also suppress 
self-intersections and, thus, effects from self-avoidance. If 
$L\ll L_p$, the semiflexible polymer approaches a rigid rod. 
For $L \gg L_p$, on the other hand, the polymer is well-described 
by a flexible polymer with an effective  segment length $\sim L_p$. 
 In the adsorption problem, both $L$  and $L_p$  also 
  compete with the correlation length $\xi$ of the adsorption transition
 and the range $\ell$ of the adsorption potential.

For the adsorption transition, polymers longer than the 
correlation length, $L\gg \xi$, can be regarded as quasi-infinite. 
Then the correlation length $\xi$ is  the characteristic maximal length 
 of desorbed segments (loops and tails) 
and diverges at the adsorption transition. 
 If the persistence length is small compared to the correlation length, 
 $L_p < \xi$, the critical exponents 
of the adsorption transition cross over to 
 those  of a flexible polymer,  if self-avoidance is taken into 
account  to those of a flexible 
self-avoiding chain. For semiflexible polymers with large  $L_p$,
   this crossover
 can only be observed very close to the adsorption 
 transition 
 \cite{Maggs1989,Kierfeld2006}. 
We will show that  the crossover from semiflexible to flexible critical 
behavior is also reflected in the  length  
distribution of the desorbed loops or tails. 
We find 
different exponents for  loop and tail distributions 
of flexible and semiflexible polymers, which give rise to  a 
different desorption behavior:
 semiflexible polymers have significantly smaller loops closed to the 
transition but  
both flexible and semiflexible polymers desorb by expanding the 
tail.  This  explains previous 
observations in  simulations \cite{Hsu2013,Welch2015}.

The critical potential strength itself
 is, however, to leading order, 
 the critical potential strength to adsorb 
single segments of size $L_p$  and, thus, 
mainly determined by semiflexible 
fluctuations on scales $<L_p$  \cite{Kierfeld2006}.
Flexible behavior only occurs if the persistence length 
becomes smaller than the adsorption potential range, $L_p < \ell$
\cite{deng10,Kampmann2013}.  
This can be qualitatively understood from the simple argument that 
adsorption of individual segments of length $L_p$ is a necessary
condition for the adsorption of the entire chain. 
The result  is in accordance with 
simulation results on the adsorption threshold, which can be well 
explained using semiflexible polymer theory \cite{Kampmann2013}
regardless of whether $\xi <L_p$ or $\xi > L_p$
as long as only $L_p\gg \ell$, i.e., the semiflexible polymer is 
sufficiently rigid that its persistence length 
 remains large compared to the 
potential range $\ell$ and loop formation of  adsorbed trains 
inside the adsorption potential remain suppressed.

 According to the theory of phase transitions, 
finite size effects will modify the adsorption behavior close 
to the transition  if 
 $L<\xi$ such that the length of desorbed loops or tails 
is limited by the finite polymer length.

For finite stiff polymers,
we find a novel rigid rod regime with qualitatively different 
finite size effects. 
In the rigid rod regime, it is essential how 
 the global orientation 
degrees of freedom are restricted by grafting or confinement.
Finite size effects then
crucially depend on  Odijk's deflection length 
\begin{equation}
  L_d \sim L_p^{1/3}\ell^{2/3},
\label{eq:Ld}
\end{equation}
 which is 
the length scale for collisions of the adsorbed 
polymer with the 
boundaries of a  potential well of width $\ell$
 \cite{Odijk1983}.
A finite stiff polymer with the  deflection length 
length $L_d$ exceeding its  length $L$,
 effectively behaves as a weakly fluctuating rigid rod. 
We will show that 
 finite size effects are then governed by the global orientation 
 of the weakly fluctuating rod if it is end-grafted 
 to the adsorbing surface.
For a stiff polymer confined by a second wall to the adsorbing surface,
also the global translation degree of freedom becomes relevant. 
For a flexible polymer, on the other hand, 
there is no preferred orientation of the 
polymer, which is thus
  irrelevant for finite size scaling.

\begin{figure*}[ht]
\centering
 \includegraphics[width=0.98\textwidth]{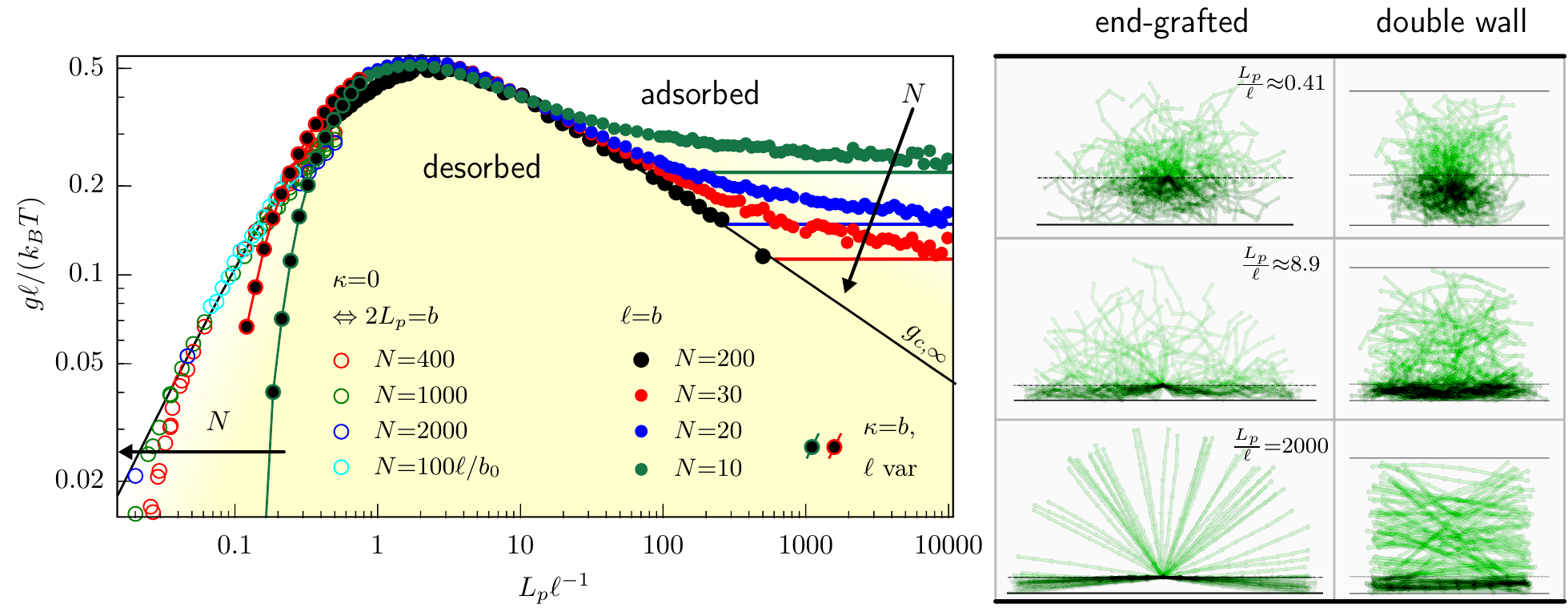}
 \caption{ 
 Simulation data for the  adsorption threshold  $g_c$ of 
     finite end-grafted polymers as a function of dimensionless 
   polymer stiffness $L_p/\ell$ 
    and three snapshots for different stiffness
     illustrating the three regimes of flexible, semiflexible and rigid rod
  adsorption. 
   We also show the corresponding snapshots for polymers confined between 
   two walls (see also Fig.\ \ref{fig:gcwall} below for the 
  simulation results on the adsorption threshold).
The
     snapshots show bearly adsorbed ($g \gtrsim g_c$, $N=10$) polymers 
     and are taken
     in two dimensions for clarity,  
    whereas all simulations in this paper were performed in
     $D{=}3$. To illustrate typical configurations, each snapshot shows
     several ($>100$) configurations. We use the maximum of the cumulant
     $C_{\rm ad}=\langle L_{\rm ad}^2 \rangle -\langle L_{\rm ad} \rangle^2 $
     to determine the adsorption threshold 
   $g_c$ numerically. For the stiffer chains (Filled colored
     circles, $L_p / \ell >1$) we vary $L_p$ and for more flexible chains
     (Colored circles with black filling, $L_p / \ell <1$) 
   we vary $\ell$. Additionally, we
     simulate chains without bending stiffness $\kappa{=}0$, where the
     persistence length is $2L_p {=}b$ according to the Kuhn length
   (colored open circles). In the stiff regime,
 horizontal lines indicate the adsorption threshold 
  of a rigid rod ($L_p \to \infty$)  from  eq.\ (\ref{eq:gcrod}).
}  
 \label{fig:gc}
\end{figure*}

The different regimes for the  adsorption threshold are summarized 
in table \ref{tab:regimes} and illustrated by simulation
snapshots in Fig.\ \ref{fig:gc}.
Understanding the influence of global degrees of freedom 
on finite size effects for stiff polymers 
 is  important  to correctly analyze 
adsorption data for finite semiflexible polymers. 
For decreasing stiffness such that 
 $L_d\ll L$,   internal deformation degrees of freedom 
dominate, and there 
is a  crossover from the novel rigid rod regime 
to  the standard finite size corrections
 from a correlation length growing beyond the contour length, 
    $\xi>L$.
From these results we can derive a procedure to analyze
finite size effects for semiflexible polymer adsorption
in simulations or
for example, in the
experimental data of Ref.\ \citenum{Welch2015} on
 the adsorption of filamentous actin.

\begin{table}
    \begin{tabular}{|c||c|c|c|}
\hline
                             &  $L_p<\ell$  &  $L_p>\ell$    &   $L_d>L$  \\
\hline\hline
     $L>\xi,L_d$ & flexible, & semiflexible,  &    n.a.\ \\
                 &   infinite &  infinite   &      \\
\hline
      $L<\xi$     & flexible,  & semiflexible,   &
                         finite   \\ 
               & fin.\ size effect  & fin.\ size effects   &
                        rigid rod   \\
\hline
\end{tabular}
\caption{Different  regimes for the adsorption threshold of 
   a continuous semiflexible polymer. 
The critical exponents should cross over 
from semiflexible to flexible (or self-avoiding flexible) 
for $\xi>L_p$ close to the desorption threshold.  
In addition,  for a semiflexible chain 
 with bond length $b$,  discretization effects  occur for $b>L_d$.}
\label{tab:regimes}
\end{table}

The paper is organized as follows. 
 In Sec.\ \ref{model},
  we introduce the theoretical and Monte-Carlo 
simulation model for the adsorption 
of  semiflexible polymers on a planar substrate.
 Then, we briefly recapitulate results on the 
adsorption threshold of infinite semiflexible polymers.
Afterwards, the loop and tail distributions
close to the adsorption transition are derived
and used  to extract 
the free energy of adsorption from 
measured adsorption data. 
We then focus on the adsorption threshold for {\it finite} semiflexible 
polymers and obtain a complete picture featuring 
a flexible, a semiflexible and a novel rigid rod regime. 
We consider different adsorption geometries, in particular, we 
 compare end-grafted semiflexible polymers 
and semiflexible polymers confined between two walls.
Finally, our results lead to a method to analyze 
adsorption data for finite semiflexible polymers
such as filamentous actin.
We  conclude with a discussion of experimental realizations. 

\section{Model and simulation}
\label{model}

We use the same semiflexible polymer model as in Ref.\ \citenum{Kampmann2013}
and start from a continuum worm-like chain 
model for a polymer contour ${\bf r}(s)$ of length $L$ 
parameterized by its arc length $s$ ($0<s<L$).
Its  energy consists of a bending energy
 ${\cal H}_b[{\bf r}(s)] = \int_0^L ds \frac{\kappa}{2} (\partial_s^2 {\br
     r})^2$ and an adsorption energy  
${\cal H}_{\rm ad}[{\bf r}(s)] = \int_0^L ds
   V(z(s))$ in an external  potential $V(z)$ that only depends 
on the distance $z$ to the absorbing surface at $z=0$,
${\cal H}= {\cal H}_b+ {\cal H}_{\rm ad}$. 
The adsorption potential is  an attractive 
 short-range square-well  part $V_a$
of range $\ell$ in front of a hard wall potential $V_{\rm wall}$ for 
a planar surface,  
 \begin{equation}
   V(z) = V_a(z)  + V_{\rm wall}(z) 
        =  \begin{cases}
                 \infty & \mbox{ for }  z < 0 \\
                   -g         & \mbox{ for } 0<z \le  \ell \\
                   0         & \mbox{ for } z >  \ell.
         \end{cases}
    \label{eq:Vz}
 \end{equation}
The potential strength  $g>0$ is an energy per length. 
Using this model, we can study adsorption both in $D=2$ and $D=3$ 
spatial dimensions. 
The persistence length of the semiflexible polymer as defined from 
the tangent-correlations of a free polymer  is 
$L_p = 2\kappa/(D-1)k_BT$ \cite{Gutjahr2006} 
 (note that the definition
 $L_p=2\kappa/k_BT$ has been used in 
 Refs.\ \citenum{kierfeld03,kierfeld04,Kierfeld2005a,Kierfeld2006,Welch2015}, 
whereas $L_p=\kappa/k_BT$ was used in Ref.\ \citenum{Kampmann2013}).
In the weak bending approximation (valid for $L,\xi <L_p$),
 we switch to  the  Monge parametrization 
 with ${\bf r}(x) = (x,y(x),z(x))$ and rewrite the energies as 
${\cal H}_b[z(x)] = \int_0^L dx \frac{\kappa}{2}  (\partial_x z)^2$ 
and ${\cal H}_{\rm ad}[z(x)] = \int_0^L dx V(z(x))$.

For the simulation, 
we discretize the semiflexible polymer into $N$ beads  connected by harmonic 
springs with a spring constant $k$ 
and a bending energy derived from the 
bending angle of three neighboring beads with a bending rigidity $\kappa$ 
 \cite{kierfeld04}. 
In the simulations, 
this semiflexible harmonic chain is a phantom chain with no  additional 
hard-core interactions between beads, i.e., there is no explicit 
self-avoidance    (as opposed to simulations in 
Ref.\ \citenum{Hsu2013}) but self-avoidance will be effectively 
fulfilled on length scales below $L_p$ because of the bending energy. 
The discretization introduces another length scale into the simulation, 
which is the rest length $b$  of the harmonic bonds resulting in 
 an  equilibrium contour length 
\begin{align}
L=(N-1)b \,.
\label{eq:L}
\end{align}
Discretization also affects the actual persistence length, which 
becomes \cite{Kleinert2004}
\begin{equation}
L_{p,{\rm dis}} = 
  \frac{b}{\ln\left( I_{D/2-1}\left(\frac{\kappa}{bk_BT}\right)/
          I_{D/2}\left(\frac{\kappa}{bk_BT}\right) \right) }
\label{eq:Lp}
\end{equation}
where $I_n(x)$ is the modified Bessel function. 
For $L_{p,{\rm dis}}/b \gtrsim 2$ the persistence length 
$L_{p,{\rm dis}}$
approaches the continuous worm-like chain result 
 $L_p = 2\kappa/(D-1)k_BT$. 
In the following, we will used the result (\ref{eq:Lp}) 
as persistence length to analyze simulation data for 
discrete semiflexible polymers.

We perform  Monte-Carlo (MC) simulations of the adsorption 
 process using a Metropolis 
 algorithm with bead displacement moves 
 of single beads or segments of beads.
 Each MC  sweep consists of $N$
  MC moves, where segments of successive  beads are moved by a random
  vector of length $v$. The MC displacement  $v$ is
  determined  before each simulation to realize  an 
     acceptance rate of  about $50\%$ (typical values are $v\simeq 0.05$).
 A typical MC simulation consists of $10^7$ sweeps.

In order to avoid that the polymer eventually diffuses away
from the adsorbing plane to infinity one has to confine the polymer 
to the adsorbing plane. 
For the analysis of finite size effects in the adsorption transition 
it will turn out to be relevant 
 how this confining mechanism  is  chosen, in particular, 
for finite stiff polymers with $L_d>L$. 
 We use an end-grafting  procedure and  
 attach one end of the polymer to the boundary 
 of the attractive potential, i.e., at $z=\ell$, in order to 
 suppress  diffusive motion of the polymer center of mass   in the 
 desorbed phase \cite{Hsu2013}.
 Confinement by end-grafting turns out to be convenient for the 
calculation of finite size effects in the rigid rod and 
stiff limit
because it eliminates global translation of 
the chain and only allows for global rotation. 
Another choice is to confine the polymer  by two hard walls
 as in Ref.\ \citenum{Welch2015}, which  allows 
for both global translation of 
the chain and global rotation  between the confining walls.

  Typical simulated polymers consist of several
   hundreds of beads. 
  In the simulation we measure  lengths in units of the bond length $b$ 
   and energies in units of $k_BT$.
 We use  values
   $k=100\,k_BT/b^2$ or
     $k=1000\,k_BT/b^2$,
     for the harmonic spring stiffness to mimic a practically 
 inextensible polymer. 
We change the persistence length $L_p$ via the stiffness $\kappa$ 
to explore finite size effects as a function of the dimensionless 
 stiffness parameter $L_p/\ell$. 

The bond length has to be sufficiently small to avoid discretization
effects.
In the semiflexible regime $L_p  >\ell$, 
a firmly adsorbed polymer 
decays into independently fluctuating  segments of the size of the 
deflection length $L_d \sim L_p^{1/3}\ell^{2/3}$ by collisions 
with the potential boundary
\cite{Odijk1983,Koster2007,Koster2008}:
thermal fluctuations 
of a weakly bent free semiflexible 
polymer of length $L$ are $\langle z^2 \rangle(L)\sim L^3/L_p$,
and the collision 
condition  $\langle z^2 \rangle(L_d)\sim \ell^2$ determines
the scaling of $L_d$.
Therefore, the simulation exhibits discretization effects 
if $b> L_d$ because these collisions form fluctuations 
within the adsorption potential layer can no longer be properly 
resolved. 
This implies a choice $(L_p/b)^{1/3}(\ell/b)^{2/3}>1$
or $(L_p/\ell)^{1/3}(\ell/b) >1$ 
to avoid discretization effects for $L_p >\ell$. 
Note that also lattice simulation such as in Ref.\ \citenum{Hsu2013}
often represent the potential range by a single layer of adhesive sites 
effectively corresponding to a contact potential $\ell \ll b$ and can, 
therefore, not resolve any fluctuations within the  adsorption potential
layer. 

In the flexible regime $L_p<\ell$ on the other hand, 
the polymer can be regarded as 
flexible polymer with an effective bond length 
of $2L_p$ also inside the potential well. 
Then discretization effects occur if $b>2L_p$  if turns 
of the polymer can no longer  be resolved properly. 
This implies a choice 
$2L_p/b>1$ to avoid discretization effects for $L_p <\ell$.

\section[Infinite polymers]{Critical potential strength of infinite polymers}

It is useful to first summarize known results for the 
adsorption transition of quasi-infinite semiflexible polymers 
($\xi<L$).
They  desorb by their 
internal configuration fluctuations, which can be envisioned
as formation of desorbed loops and tails. 
The maximally accessible  desorbed loop and tails size is limited
by the correlation length $\xi$. All fluctuations on scales 
$<L_p$ are governed by bending energy of a semiflexible polymer, 
whereas fluctuations on scales $\gg L_p$ can be regarded as 
fluctuations of a flexible polymer with an effective bond length 
of $2L_p$.

\subsection{Semiflexible  regime}

Infinite semiflexible polymers  desorb by 
internal configuration fluctuations. 
If the persistence length exceeds the correlation length, $L_p \gtrsim \xi$,
the desorbed ``loops'' are actually oriented  segments without turns. 
If  the persistence also exceeds 
 the potential range $\ell$, $L_p \gtrsim \ell$, the  adsorbed 
tail segments cannot perform turns within the potential range.
Then 
 we are in the stiff or semiflexible limit of adsorption, where  
the  critical potential strength is 
\begin{equation}
 g_{c,SF}  = c_{SF} \, \frac{k_BT}{\ell} \left(\frac{L_p}{\ell}\right)^{-1/3}  
     ~~~(L_p \gtrsim \ell).
\label{gcSF}
\end{equation}
The  exact exponent $1/3$ occurring in eq.\ (\ref{gcSF}) 
has been obtained using different approaches:
This result has been derived  in 
Ref.\ \citenum{Gompper1990} via the necklace model approach \cite{Fisher1984}.
It has also been obtained
explicitly in  Refs.\ 
\citenum{Netz1999,Semenov2002,kierfeld03}
by scaling arguments 
\cite{Netz1999} or analytical transfer matrix calculations 
\cite{Semenov2002,kierfeld03}. 
Refs.\ \citenum{Kierfeld2006,Kampmann2013} 
(see also Supplemental Material of both Refs.) contain a more detailed
account of analytical transfer matrix calculations of this result.
In Ref.\ \citenum{Kampmann2013} (and Supplemental Material of 
\citenum{Kampmann2013}),
there are also results for the numerical prefactor 
\begin{equation}
c_{SF} = 2^{-2/3} 3^{-1/3}\Gamma(1/3) (D-1)^{-1/3} \simeq 0.93 
\left(\frac{2}{D-1}\right)^{1/3}
\label{eq:cSF}
\end{equation}
in eq.\ (\ref{gcSF}) (remember the different definition $L_p = \kappa/k_BT$ 
   used in Ref.\ \citenum{Kampmann2013}).
In Ref.\ \citenum{Welch2015} the exponent $1/3$ in  
the result   (\ref{gcSF}) has been ``rediscovered'' ignoring all of these
previous explanations. 

The  scaling with an exponent $1/3$ in the result   (\ref{gcSF}) for the 
critical potential strength is directly related to a 
corresponding scaling dependence of Odijk's deflection 
length $L_d \sim L_p^{1/3}\ell^{2/3}$.
This relation 
 is  established by a standard statistical mechanics
argument:
Confinement to the potential well costs entropy of the order
of $1k_B$ per collision with the potential boundary, 
which gives a free energy cost per length
 $\Delta f = T \Delta s = k_BT/L_d$.
Balancing this with the energy gain $g$ per length gives
the scaling of the adsorption 
 threshold ad $g_c \sim k_BT/L_d$ resulting 
in eq.\  (\ref{gcSF}). 

As a consequence, lattice simulations that cannot resolve
fluctuations within the potential layer due to discretization effects,
such as in Ref.\ \citenum{Hsu2013},
will find a different scaling behavior of the 
adsorption threshold, which is  dominated by discretization effects.
A semiflexible lattice polymer will change lattice direction on its
persistence length $L_p$ on average. 
If the adsorption layer is a 
 single layer of adhesive sites, confinement to this layer suppresses
a finite fraction of configuration on the lattice
every persistence length $L_p$, which 
leads to an 
  entropy cost $T\Delta s \sim k_BT/L_p$ per length and, thus, 
to an adsorption threshold $g_c \sim k_BT/L_p$ as it was observed in 
Ref.\ \citenum{Hsu2013}.

In Ref.\ \citenum{Kierfeld2006} it has  been pointed out
that the result (\ref{gcSF}) for the critical potential strength
remains valid also if $L_p <\xi$ because it represents the 
critical potential strength to adsorb 
single segments of size $L_p$.
In fact, the result (\ref{gcSF}) holds  as long as these segments are 
larger than the potential range $\ell$,  $L_p\gg \ell$
 \cite{deng10,Kampmann2013}. 
Then the semiflexible polymer is 
sufficiently rigid that loop formation of  adsorbed trains 
inside the adsorption potential remain suppressed.

\subsection{Flexible regime}

The semiflexible result
eq.\  (\ref{gcSF}) is  applicable if $L_p$ is 
larger than the potential range $\ell$ \cite{deng10,Kampmann2013}.
For $L_p\lesssim \ell$, the polymer is in the flexible regime, 
where it can perform turns within the potential range. 
As it has been observed and analyzed 
in Refs.\  \citenum{deng10,Kampmann2013},
there is a maximum in the critical potential $g_c\ell$ 
 for $L_p/\ell \sim 1$, such that 
adsorption becomes easier again in the flexible limit $L_p \lesssim \ell$,
where 
\begin{equation}
 g_{c,F}  = c_F \, \frac{k_BT}{\ell} \frac{L_p}{\ell}
  ~~~(L_p \lesssim \ell)
\label{gcF}
\end{equation}
is found with $c_F  = 2\pi^2/4D(D-1)$ in $D$ spatial dimensions
in the absence of self-avoidance.
The result (\ref{gcF}) is the standard result 
$g_{c,F} \sim k_BT b/\ell^2$  for a flexible ideal
 chain \cite{degennes}
with effective bond length $b=2L_p$. 
Again, lattice simulations can only find such a scaling behavior 
if the potential range is represented by several 
lattice spacings \cite{Klushin2013}.

\subsection{Crossover between regimes}

In Ref.\ \citenum{Kampmann2013}, the interpolation formula 
\begin{align}
   \frac{g_c\ell}{k_BT} &=  I\left(\frac{L_p}{\ell}\right)
 \label{eq:I}\\
  I(x) &= c_1x (1+c_2 x^{4/3}+c_3 x^{2/3} )^{-1}
\label{eq:fit1}
 \end{align} 
has been derived,
 which describes
both stiff and flexible limits and 
 contains three free fit parameters $c_1$, $c_2$, and $c_3$.
The choices  $c_1=c_F$  
 and $c_2  =  c_F/c_{SF}$  
  reproduce the analytically known  flexible and semiflexible
 limits,  
 and the remaining  parameter $c_3$ allows to vary the position 
 of the maximum to fit simulation data. 
In $D=3$ we find best fits  of our MC simulation data 
for parameter values as given in table \ref{tab:fits};
these values slightly differ from results in Ref.\ \citenum{Kampmann2013}
because we include discretization effects properly by using the 
persistence length eq.\ (\ref{eq:Lp}); this improves the fit to 
the simulation data for small $L_p/b \lesssim 2$.

\begin{table}
    \begin{tabular}{|c||c|c|c|}
\hline
data set  & $c_1 /c_{F}$    & $c_2 c_{SF} /c_1$   & $c_3$  \\ 
\hline
\hline
theory (D=3) &  $1 $       & $1$  &  free  \\
cumulant      & $1.50{\pm} 0.05$  & $1.03 {\pm} 0.01$     & $0.52 {\pm} 0.06$
\\
finite size   & $1.08{\pm} 0.05$  & $1.04 {\pm} 0.02$     & $0.74 {\pm} 0.1$ \\
\hline
\hline
theory (D=2) &  $1 $       & $1$  &  free  \\
cumulant      & $1.05{\pm} 0.05$  & $1.03 {\pm} 0.01$     & $0.00 {\pm} 0.06$
\\
finite size   & $0.83{\pm} 0.05$  & $1.05 {\pm} 0.02$     & $0.52 {\pm} 0.1$ \\
\hline
\end{tabular}
\caption{Simulation results for the fit 
  parameters  $c_1$, $c_2$, and $c_3$  for the interpolation
  function $I(x)$ from eq.\  (\ref{eq:fit1}) 
in comparison 
  with  theoretical expectations for $D=3$ and $D=2$. 
The adsorption threshold is determined by the cumulant 
method  (maximal adsorption energy fluctuations) or 
finite size scaling. 
}
 \label{tab:fits}
\end{table}

The MC data in 
Fig.\ \ref{fig:gc} confirms that long polymers approach these
results for infinite polymers. The data also shows 
pronounced finite size effects and a third adsorption regime in the 
stiff rod limit, which we will address below.

\section{Theory of loops and tails}

\subsection{Return exponents}

Before discussing the adsorption of finite polymers it is useful
to understand how the diverging correlation length $\xi$
governs the length distribution of desorbed loops and tails 
 of a semiflexible or flexible polymer. 
The length distributions of loops and tails has been  examined
recently in simulations \cite{Hsu2013,Welch2015}, where it was found 
that stiffer semiflexible polymers tend to desorb from their end 
by increasing tail lengths.  

We want to explain the observations in Refs.\ \citenum{Hsu2013,Welch2015}
 using results from 
transfer matrix approaches and  necklace models based 
on the grand canonical partition sum 
\cite{Birshtein1979,Bhattacharya2009,Klushin2013} (see 
Ref.\ \citenum{Fisher1984} for a review on necklace models). 
The relevant quantity characterizing the size distribution of loops
is the return or loop exponent $\chi$, which determines the probability 
$p_{\rm  return}\sim L^{-\chi}$ 
for a {\it free} polymer starting in the attractive region of  
the potential well to return 
 to this region for the first time as a function of 
its (projected) length $L$ in the large $L$ limit.
The first return or loop exponent $\chi$ determines 
the critical properties of the adsorption transition. 
This can be seen  in the necklace model approach, where 
 the grand canonical partition sum is written as alternating series 
of bound (train) segments and loop segments
For polymer adsorption, we need the exponent $\chi$ for 
 first  returns 
to the attractive potential well at $0<z<\ell$ in front 
of a hard wall.

In the following we
distinguish between unconditioned returns,
for which we use an exponent $\tilde{\chi}$,
and  first returns, for which we use the 
exponent $\chi$. 
 There is the general relation 
\begin{equation}
\chi= \max(2-\tilde{\chi},\tilde{\chi})
\label{chitildechi}
\end{equation}
between unconditional and first returns, 
which holds both for flexible and  semiflexible chains 
and which  can be derived, for example, 
using generating functions
and  the necklace 
representation \cite{Fisher1984}.

For a flexible polymer without self-avoidance, 
the unconditioned return exponent 
in the absence of a wall is  the standard 
result  $\tilde{\chi}_{\rm RW}=D/2$ for the return of Gaussian
 chains or random walks in $D$ dimensions to the 
origin (starting point).
 For the adsorption of Gaussian chains 
this result is applied to a single dimension ($D=1$), namely 
the $z$-coordinate of the polymer contour, which has to 
return to the potential well at $0<z<\ell$,
and with the arclength $s$  as time-like coordinate of the random walk:
 $p_{\rm  return}\sim L^{-1/2}$ is the probability that the $z$ coordinate 
returns to the adsorbing state $z\approx 0$ after length $s=L$, i.e., 
$\tilde{\chi}_{F,0}=1/2$ in the absence of the hard wall at $z=0$.
In front of a hard wall, introduction of an image walker leads 
to a return exponent $\tilde{\chi}_{F}=3/2$ for flexible polymers or random 
walks. This also follows from the observation that for flexible chains 
returns in front of a hard wall are equivalent to first returns
in the absence of a hard wall 
such that, according to (\ref{chitildechi}), 
$\chi_F = \tilde{\chi}_F  = 2-\tilde{\chi}_{F,0} = 3/2$.

The unconditioned return exponent for the weakly bent semiflexible 
phantom polymer to the adsorption potential  in the presence of a 
hard wall 
is  $\tilde{\chi}_{SF}= 5/2$ for  return with orientation
parallel to the wall. This is a non-trivial result 
which derives 
from an Ornstein-Uhlenbeck process and  models 
 a random walker  with inertia, i.e., with 
random acceleration in 1 spatial dimension \cite{burkhardt}
(for semiflexible polymers
returns in the presence of a hard wall are not 
equivalent to first returns in the absence of a wall 
because of additional restrictions on tangent 
vectors from the hard wall).  
For the adsorption problem
this result is applied to the $z$-coordinate of the polymer contour
and with the projected 
length   $x$ along the preferred orientation of the weakly 
bent semiflexible polymer as time-like coordinate of the 
inertial random walk:
 $p_{\rm  return}\sim L^{-5/2}$ is the probability that the $z$-coordinate 
returns to the adsorbing state $z\approx 0$ in parallel 
orientation $\partial_x z=0$ after the projected 
length $x=L$ and in the presence of a hard wall at $z=0$. 
 The corresponding exponent 
 for return 
irrespective of orientation is
 $\tilde{\chi}_{SF}= 2$  as integration over tangents always 
gives an additional factor $L^{1/2}$ reducing the exponent
\cite{Kierfeld2005a}.

Alternatively, the exponents  $\tilde{\chi}_{F}= 3/2$ and 
$\tilde{\chi}_{SF}= 5/2$
follow from the exponent relation $\tilde{\chi}= 1+ \nu_R$ 
\cite{Lipowsky1995}, which holds in $D=1$ dimensions
(i.e., for returns in the $z$-coordinate) and 
in which  $\nu_R$ is the exponent characterizing the 
 end-to-end distance 
$\langle R^2 \rangle \sim L^{2\nu_R}$ and, thus, also 
the roughness 
in the $z$-coordinate $\langle z^2 \rangle \sim L^{2\nu_R}$.
For semiflexible polymers $\nu_{R,SF}=3/2$ resulting in $\tilde{\chi}_{SF}= 5/2$;
for flexible phantom polymers $\nu_{R,F}=1/2$ resulting in 
$\tilde{\chi}_{F}=3/2$.

Using the relation (\ref{chitildechi}) between unconditioned
and first returns we see 
that for adsorption in front of a hard wall, as considered 
here, return and first return (loop) exponents are identical and 
\begin{align}
   \chi_F &= \tilde{\chi}_F = 3/2, \nonumber\\
  \chi_{SF} &= \tilde{\chi}_{SF} = 2.
\label{chi}
\end{align}

Loop or first return exponents are also known for self-avoiding chains,
where 
\begin{equation}
\chi_{{\rm SAW}} = 
\begin{cases} 
   1-\gamma_{11} \simeq 1.39 & \mbox{in}~D=3\\
  19/16 = 1.1875    & \mbox{in}~D=2
\end{cases}
\label{chiSAW}
\end{equation}
holds for  returns of self-avoiding chains 
 to a short-range  adsorption potential in front of a hard wall
\cite{Grassberger2005,Bhattacharya2009,vanderzande1998}.
Note that these result are close but not identical to 
the naive estimate 
 $p_{\rm  return}\sim L^{-\nu_R}$ or 
$\tilde{\chi}_{{\rm SAW},0} = \nu_{R,{\rm SAW}}$ using 
the exponent $\nu_R$ of the  end-to-end distance 
$\langle R^2 \rangle \sim L^{2\nu_R}$ and assuming that return 
in the $z$-coordinate is simply governed by the extension 
of the polymer, 
$p_{\rm return} \sim 1/\langle R^2 \rangle^{1/2}$.
Assuming also that 
 returns in front of a hard wall are equivalent to first returns 
 in the absence of a hard wall, we have
$\tilde{\chi}_{\rm SAW}= 2- \tilde{\chi}_{{\rm SAW},0} \simeq 2-\nu_{R,{\rm SAW}}$
 according to (\ref{chitildechi}).
The Flory estimate $\nu_{R,{\rm SAW}} = 3/(D+2)$ for self-avoiding polymers
gives  $\nu_{R,{\rm SAW}} = 3/5$ in $D=3$, more exact 
 field-theoretical calculations  give 
 $\nu_{R,{\rm SAW}} \simeq 0.588$ \cite{LeGuillou1980};
$\nu_{R,{\rm SAW}}=3/4$ is exact in $D=2$ \cite{vanderzande1998}. 
The reason for deviations from these naive estimates in eq.\
(\ref{chiSAW}) are 
additional correlations between the $z$-component of the walk 
and components parallel to the surface by self-avoidance. 
Interestingly, 
in Ref.\ \citenum{Bhattacharya2009} it was shown that 
additional self-avoidance interactions between different adsorption 
loops exactly restore the naive result
such that 
\begin{equation}
\tilde{\chi}_{\rm SAW}= 2-\nu_{R,{\rm SAW}} = 
\begin{cases} 
   \simeq 1.41 & \mbox{in}~D=3\\
    = 1.25    & \mbox{in}~D=2.
\end{cases}
\label{chiSAW2}
\end{equation}
According to  (\ref{chitildechi}),
also for self-avoiding chains, 
 return and first return exponents are identical in front of a hard wall,
i.e., 
$\chi_{\rm SAW} = \tilde{\chi}_{\rm SAW}= 2- \tilde{\chi}_{{\rm SAW},0}$. 

We note that, comparing loop exponents for flexible, self-avoiding 
and semiflexible polymers, we find 
$\chi_{\rm SAW}<\chi_{F}<\chi_{SF}$ (see (\ref{chi}) and (\ref{chiSAW2})), 
i.e., the exponent 
is smallest for the self-avoiding chain, although
the exponents $\nu_R$ for the end-to-end distance  
are ordered differently, 
$\nu_{R,F}< \nu_{R, {\rm SAW}} <\nu_{R,SF}$ (see (\ref{nu}).

\subsection{Order of adsorption transition}

 Necklace model \cite{Fisher1984} and transfer matrix approaches
 \cite{Kierfeld2005a}  independently show that 
the return exponent $\chi$  determines all other critical exponents
of the adsorption transition. 
In particular, 
there is a  relation between the first return 
exponent $\chi$ and the correlation length exponent $\nu$ of the adsorption 
transition ($\xi \propto |g-g_c|^{-\nu}$,) which is 
\begin{equation}
  1/\nu = \min(\chi-1,1)
\label{nuchi}
\end{equation} (correcting an error in Ref.\ \citenum{Kierfeld2005a}). 
This relation  gives 
\begin{align}
 \nu_{F}&=2,\nonumber\\
 \nu_{\rm SAW}& = \frac{1}{1-\nu_R}  =
\begin{cases} 
   \simeq 2.43 & \mbox{in}~D=3\\
    = 4  & \mbox{in}~D=2.
\end{cases}
\nonumber\\
  \nu_{SF} &= 1+{\rm log}
\label{nu}
\end{align}
from (\ref{chi}) \cite{kierfeld03,Kierfeld2005a,Kierfeld2006}
and (\ref{chiSAW}).

For the adsorption transition, transfer matrix  theory 
shows that the exponent $\nu$ 
is identical to the free energy exponent describing the 
singular part of the adsorption free energy density 
$|f_{\rm ad}| \sim k_BT/\xi \propto  |g-g_c|^{\nu}$
\cite{Fisher1984,kierfeld03} (i.e.\ hyperscaling holds).
Thus, 
also the nature of the phase transition 
is entirely determined by 
the first return or loop exponent $\chi$:
If $\chi<1$ the polymer is always bound for arbitrary weak potential;
this case  is  not possible for a first return 
probability as immediately 
follows from the above relation (\ref{chitildechi}). 
If $\chi>1$ a threshold potential strength is necessary for adsorption.
For $1<\chi<2$ or $\nu = 1/(\chi-1)>1$ the transition is continuous. 
For $\chi >2$ or $\nu =1$ the transition becomes discontinuous. 
  So $\chi=2$ and $\nu=1$ marks the boundary 
between a discontinuous and continuous transition.
The semiflexible polymer with $\chi_{SF}=2$ is  at this 
boundary and is 
weakly second order with $\nu_{SF}=1+{\rm log}$ as 
closer inspection shows \cite{kierfeld03}.
Both flexible and self-avoiding chains have  $1<\chi_{F},\chi_{\rm SAW}<2$
such that adsorption is continuous.

 If the persistence length is small compared to the correlation length, 
 $L_p < \xi$, there is a crossover in the   critical properties, i.e.,
the critical exponents 
 of the adsorption transition, to 
 those  of a flexible polymer, eventually 
a flexible self-avoiding chain. 
Because the correlation length $\xi\sim k_BT/|f_{\rm ad}|$ 
ultimately diverges at the transition,
the critical properties observable right at  the adsorption transition 
should {\em always} be those of flexible chains with $\nu_{F}=2$ or 
$\nu_{\rm SAW}=1/(1-\nu_R)\simeq 2.43$ in the presence of 
self-avoidance. 
For stiff polymers with large $L_p$ this critical behavior is,
however, only observable for $|g-g_c|< k_BT/L_p$ (because 
of $\nu_{SF} \approx 1$), i.e., very close to the critical point
\cite{Kierfeld2006}. 
As long as $|g-g_c|> k_BT/L_p$, the transition should have 
{\em apparent} critical exponents from the semiflexible regime
\cite{Kampmann2013}. 
Only in the rigid rod limit $L_p\to \infty$, a truly 
discontinuous transition as predicted from the semiflexible 
criticality should be observable. This crossover in critical 
properties often causes confusion, see for example a recent discussion 
in Ref.\ \citenum{Hsu2013}.

\subsection{Finite size scaling results for critical properties}

In principle, 
the correlation length or free energy 
exponent $\nu$ can be determined by finite size scaling
of the adsorbed length  $\langle L_{\rm ad} \rangle 
= -\langle {\cal H}_{\rm ad} \rangle/g$ or its 
second  cumulant  $C_{\rm ad} \equiv \langle L_{\rm ad}^2 \rangle -
\langle L_{\rm ad} \rangle^2$.
Because $C_{\rm ad} =  -k_BT L \frac{\partial^2 f_{\rm ad}}{\partial g^2}$
and $|f_{\rm ad}| \sim k_BT/\xi \propto  |g-g_c|^{\nu}$, we have
$C_{\rm ad}\propto L |g-g_c|^{\nu-2}\propto L \xi^{-1+2/\nu}$,
which results in a finite size scaling
\begin{equation}
   C_{\rm ad} = L^{2/\nu} f((g-g_c)L^{1/\nu}) 
\label{eq:Cad}
\end{equation}
for $g>g_c$ with a scaling function $f(x)$.
This type of standard finite size scaling applies 
if the polymers are long enough to avoid the crossover to 
the additional rigid rod finite size corrections which will be discussed
below in Sec.\ \ref{sec_finite}.

\begin{figure}[htb]
\centering
 \includegraphics[width=0.46\textwidth]{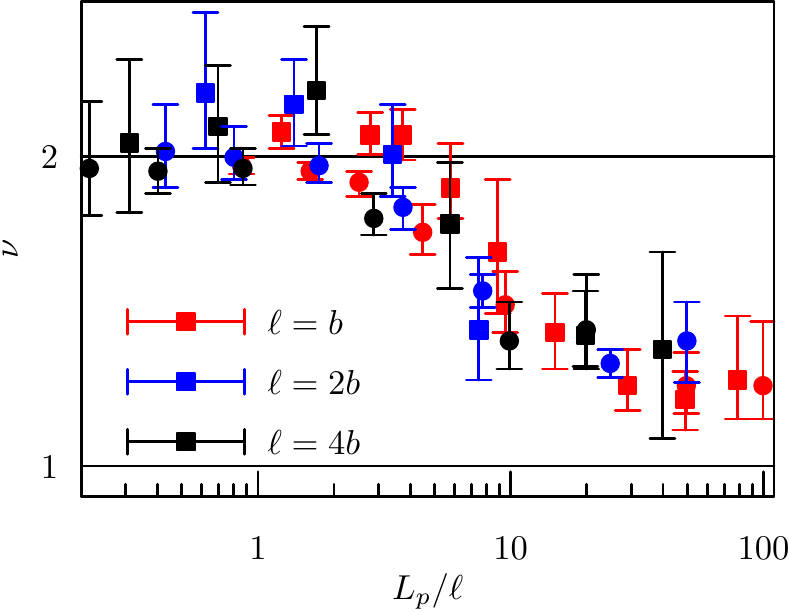}
 \caption{
Finite size scaling results for the exponent $\nu$ in $D=2$ (squares)
and $D=3$ (circles) and for $\ell=b$ (red), $\ell=2b$ (blue),
$\ell = 4b$ (black) as a function of the stiffness parameter 
$L_p/\ell$. Polymer lengths used in finite size scaling are 
$N=50-800$, i.e., polymers are long enough to avoid the crossover 
 to the rigid rod regime discussed in Sec.\ \ref{sec_finite}.  
}
 \label{fig:nu}
\end{figure}

An  analysis of simulation data (see also \citenum{Kampmann2013} and 
Supplemental Material of \citenum{Kampmann2013} for 
more details)  shows that the critical exponent measured by this approach 
indeed changes from $\nu\approx \nu_F=2$ for $L_p/\ell <1$ 
to $\nu \approx \nu_{SF}=1$  for $L_p/\ell >10$, see Fig.\ \ref{fig:nu}. 
The contour lengths and potential ranges 
used for finite size scaling in Fig.\ \ref{fig:nu} are
$L/b=50-800$ and $\ell/b=1-4$, respectively 
 (therefore, for $L_p/\ell <100$ the deflection length 
$L_d\sim L_p^{1/3}\ell^{2/3}$ is small compared to the contour length,  
$L_d< 20b <L$, such that the 
rigid rod limit and corrections from global rotation degrees 
of freedom to be discussed below can be neglected). 
For infinite semiflexible polymers, the crossover
to the critical properties of flexible polymer  adsorption
should happen for $\xi >L_p$.
For finite semiflexible polymers this crossover 
should remain  unobservable if $L\lesssim L_p<\xi$ because the finite 
polymer cannot explore the relevant fluctuation wavelengths $>L_p$.
Fig.\ \ref{fig:nu} shows, however, 
 that the semiflexible criticality for  $L_p/\ell >10$ 
remains observable even if much longer 
contour lengths up to $L/L_p=80$ are use for finite size scaling.
This suggests that $L$ has to be substantially larger than $L_p$ 
 in order  to observe the flexible polymer criticality.

\subsection{Loop size distribution}

The loop size distribution 
takes the form 
\begin{equation}
 p_{\rm loop}(l) \sim l^{-\chi} e^{-l/\xi}.
\label{ploop}
\end{equation}
with the loop exponent $\chi$ from eq.\ (\ref{chi}). 
This loop size distribution as derived from transfer matrix or 
necklace models only holds 
for large loops $l$ exceeding 
any microscopic scales set by, e.g., the potential range $\ell$ such 
as  the deflection 
length $L_d \sim L_p^{1/3}\ell^{2/3}$, which acts like an effective 
 segment length for a bound semiflexible polymer.
Loops sizes are cut off exponentially at the 
correlation length $\xi$.
The shape of the  loop size
distribution (\ref{ploop})
including the exponential cut off by the correlation length 
follows from transfer matrix treatments of the problem
\cite{kierfeld03,Kierfeld2005a}. 
For a polymer segment of length $l$ starting at 
$z=0$ and ending at $z=0$, the number of loop configurations 
(not touching the potential in between) is given by the 
restricted partition sum $Z_0(z,z',l)$ of a free polymer 
as $Z_0(0,0,L) \sim  l^{-\chi}$. 
For an adsorbed polymer, the transfer matrix treatment shows the relation 
 $|f_{\rm ad}| = k_BT/\xi$  between the free energy per length (relative to
$f_0=0$ for the free polymer, $f_{\rm ad}<0$ in the adsorbed phase) 
 and  the correlation length $\xi$ 
of the transition \cite{Fisher1984,kierfeld03,Kierfeld2005a}. 
The total partition sum of the adsorbed segment is, thus, 
$Z_{ad}(l)= \exp(-f_{\rm ad}l/k_BT)$. The 
probability to find a loop of size $l$ is 
\begin{equation}
 p_{\rm loop}(l) = \frac{Z_0(0,0,L)}{Z_{ad}(l)},
\end{equation}
which is of the form (\ref{ploop}) with $\xi = k_BT/|f_{\rm ad}|$ 
($f_{\rm ad}<0$).

From the larger return exponent $\chi_{SF}=2$ as compared to  $\chi_{F}=3/2$ 
or $\chi_{\rm SAW}\simeq 1.412$ (in $d=3$), see eq.\ (\ref{chi}), 
 it follows that the loop 
size distribution shifts its weight to smaller sizes
for stiffer polymers. 
For $\chi\le 2$, the mean loop size $\langle l \rangle_{\rm loop} = 
  (\int_0^\infty dl l p_{\rm loop}(l))/(\int_0^\infty dl  p_{\rm loop}(l))$ 
diverges with 
$\xi$ as $\langle l \rangle_{\rm loop} \sim \xi^{2-\chi}$ for  $1<\chi<2$
and $\langle l \rangle_{\rm loop} \sim \ln\xi$ for $\chi=2$. 
For finite polymers with $L<\xi$ loop sizes are cut off at the polymer 
length $L$ and we find 
$\langle l \rangle_{\rm loop} \sim L^{2-\chi}$ for  $\chi<2$
and $\langle l \rangle_{\rm loop} \sim \ln L$ for $\chi=2$ accordingly. 
We conclude that loop sizes are much smaller for semiflexible 
polymers, where we only find a weak log-divergence close to 
desorption ($\chi_{SF}=2$) as compared to  flexible polymers 
 ($\chi_{F}=3/2$), where $\langle l \rangle_{\rm loop} \sim L^{1/2}$
(for $L< \xi$). 
This has also been observed in Ref.\ \citenum{Welch2015}
in simulations. 

For increasing loop size $l$ we expect to see a crossover from 
a semiflexible behavior with $\chi_{SF}=2$ for loops $l<L_p$ to 
a phantom  or self-avoiding flexible behavior for loops $l>L_p$.
The result  $\chi_{\rm SAW}\simeq 1.412$ for the self-avoiding walk 
is in qualitative agreement with an exponent  $\chi_{\rm SAW}\simeq 1.3$
observed in simulations in 
Ref.\ \citenum{Hsu2013} for the loop distribution 
of very long self-avoiding semiflexible chains $l>L_p$. 
In order to probe the regime $l>L_p$ for $l<\xi,L$, 
it is necessary to have $L,\xi \gg L_p$,
which is the regime investigated  in Ref.\ \citenum{Hsu2013}.
Experimentally, the flexible or self-avoiding regimes 
 should be accessible, for example, 
 for long DNA-strands with $ L\gg L_p \simeq 50{\rm nm}$
close to their adsorption transition. 
For polymers with $L\lesssim L_p$ as for F-actin as 
investigated in Ref.\ \citenum{Welch2015} or also for microtubules 
we rather expect to see semiflexible behavior of adsorption loops.

The exponents $\chi_F=3/2$ for flexible polymers and 
$\chi_{SF} = 2$ for semiflexible polymers from 
 eq.\ (\ref{chi}) are confirmed by our  MC simulations
of long polymers without self-avoidance
as shown in Fig.\ \ref{fig:loop}.

\subsection{Tail size distribution}

The corresponding length distribution of desorbed tails 
follows from 
interpreting a tail as the beginning of a loop. 
Therefore the tail distribution follows from integrating 
over all possible completions to a loop of size $s>l$, 
\begin{align}
  p_{\rm tail}(l)  &= \int_l^\infty  ds p_{\rm loop}(s)
\nonumber\\
 p_{\rm tail}(l)  &\sim 
   \left\{ \begin{array}{ll} 
   l^{-(\chi-1)} \exp(-l/\xi), &\chi>1\\
   l^{-\chi} \exp(-l/\xi),  &\chi<1
   \end{array}
\right. 
\label{ptail}
\end{align}
resulting in a tail exponent $\chi_F-1=1/2$ for flexible polymers,
$\chi_{\rm SAW}-1 \simeq 0.412$ for self-avoiding flexible polymers (in $d=3$),  
and $\chi_{SF}-1=1$ for semiflexible polymers. 
The reduction of the exponent by one shows that tails 
are always much larger than loops at the adsorption 
transition where $\xi\to \infty$. 
Both become limited by $\xi$  in the adsorbed phase. 
For the  mean tail size $\langle l \rangle_{\rm tail} = 
  (\int_0^\infty dl l p_{\rm tail}(l))/(\int_0^\infty dl  p_{\rm tail}(l))$,
we find $\langle l \rangle_{\rm tail} \sim \xi$ for  $1<\chi<2$
and $\langle l \rangle_{\rm tail} \sim  \xi/\ln\xi$ for $\chi=2$. 
For finite polymers with $L< \xi$, the the cutoff $\xi$ is replaced 
by the length $L$ as for loops. 
Both for flexible and semiflexible polymers of length $L$, tails are 
diverging as $\langle l \rangle_{\rm tail} \sim L$ with logarithmic
corrections 
for semiflexible polymers.   

Comparing loop and tail sizes, we conclude that both flexible 
and semiflexible 
polymers desorb by expanding tails over the whole polymer. 
Loop sizes at the transition are, however, significantly larger for flexible 
polymers. This explains the simulation results  
 in Refs.\ \citenum{Hsu2013,Welch2015}.

\begin{figure}[htb]
\centering
 \includegraphics[width=0.46\textwidth]{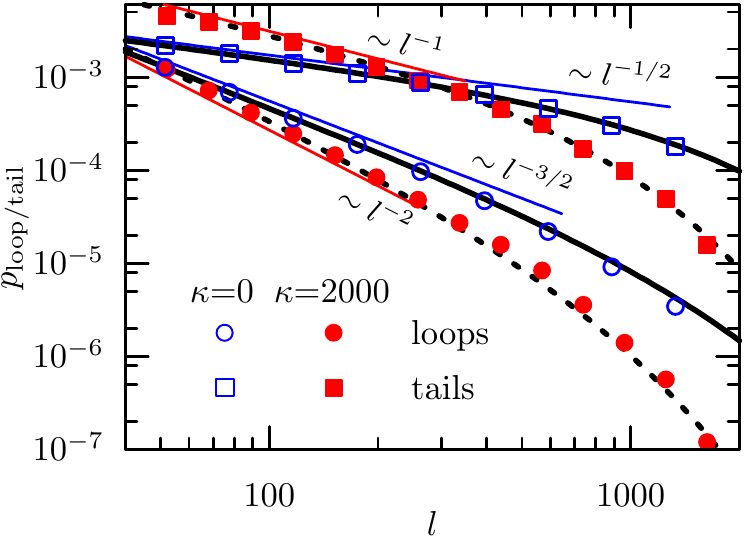}
 \caption{
Loop and tail size distributions for flexible ($\kappa{=}0$) and
   stiff polymers ($\kappa=2000$). The expected power law behavior
   (see eqs.\ \eqref{ploop} and \eqref{ptail}) on intermediate 
   length scales $b\ll l < \xi$
   is marked as colored lines.  We obtain return exponents 
 $\chi_{F}=1.487(2)$, $\chi_{SF}=1.85(1)$ and  correlation lengths
   $\xi_{F}/b=1471(65)$ and $\xi_{SF}/b=589(21)$ via an iterative
   fit scheme (see text). 
 }
 \label{fig:loop}
\end{figure}

As for loops, we also expect 
for increasing tail sizes $l$  to see a crossover from 
a semiflexible behavior with an exponent $\chi_{SF}-1=1$ for loops $l<L_p$ to 
a  flexible behavior for loops $l>L_p$ 
with $\chi_F-1=1/2$ in the absence of self-avoidance 
and to $\chi_{\rm SAW}-1 = 1-\nu_{R,{\rm SAW}}$ for self-avoiding flexible 
polymers.

Fig.\ \ref{fig:loop} shows our MC simulation results for the 
loop and tail  distributions of long flexible polymers ($\kappa=0$)
and  semiflexible  polymers ($L_p/b\simeq 4000$, simulations in  $D=2$) 
without self-avoidance and 
 with $N= L/b+1=2000$ beads 
close to the transition such that the correlation length 
$\xi \lesssim L$ and loops and tails 
occur up to large sizes. 
We find $\chi_{F}=1.487(2)$ and  $\chi_{SF}=1.85(1)$
in qualitative agreement with the theoretical result eq.\ (\ref{chi}). 
We determine the exponents $\chi$ and 
and the corresponding correlations lengths $\xi$ 
from fitting  loop and tail distributions simultaneously and 
iteratively using eqs.\ (\ref{ploop}) and (\ref{ptail}):
We first fit the tail distribution with $\xi$ as fit parameter 
at fixed  $\chi$, then the 
 loop distribution with fixed $\xi$ and with $\chi$ as a fit parameter
until  $\xi$ and $\chi$ converge. We omit small loop and tail 
size $l/b<40$ for the fits (the potential range is $\ell=0.1b$ such that 
$L_d \simeq 3.4$, i.e., only loops $l\gg L_d$ are considered).

\subsection{Analyzing loop and tail distributions} 

The loop and tail distributions not only give  insight into the 
desorption process 
and the differences between flexible and stiff  polymers.
The exponential cutoff $p_{\rm train}(l) \propto \exp(-l/\xi)$ 
can also be used to determine the correlation length $\xi$ and, thus, 
the free energy of adsorption via the relation
$|f_{\rm ad}| = k_BT/\xi$.
The free energy of adsorption is otherwise difficult to determine 
experimentally. 

Both  loop and tail distributions (\ref{ploop}) hold 
for loops $l$ longer than 
  the  deflection 
length $L_d \sim L_p^{1/3}\ell^{2/3}$, which acts like an effective 
 segment length for the bound trains  in the model.
Therefore, we propose to fit loop and 
tail distributions from simulations or
experiments using eqs.\ (\ref{ploop}) and (\ref{ptail}) 
for $l>L_d$  using $\chi$ and 
 $\xi=k_BT/|f_{\rm ad}|$ as  fit parameters.
For simulations we  performed such fits in Fig.\ \ref{fig:loop}.

\begin{figure}[htb]
\centering
 \includegraphics[width=0.46\textwidth]{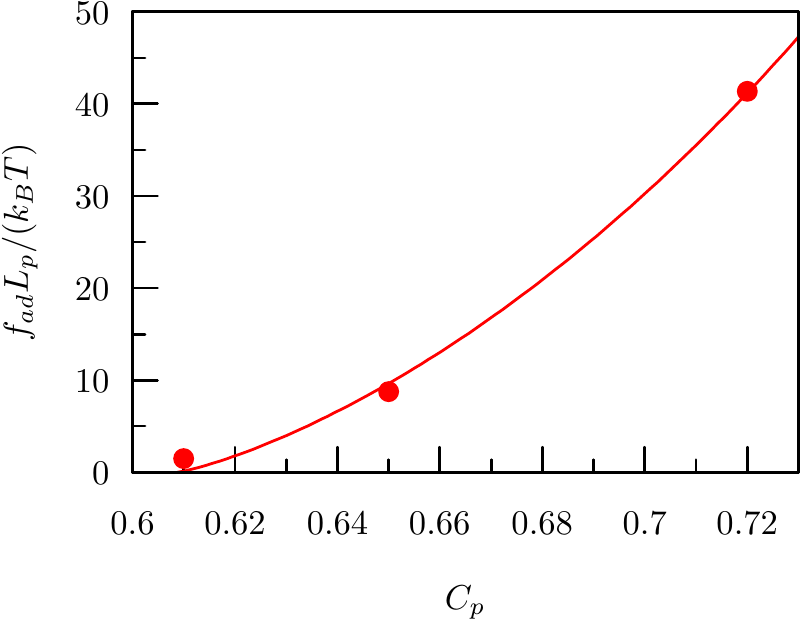}
 \caption{
Free energies of adsorption $|f_{\rm ad}|$ (in units of $k_BT/L_p$) 
from the tail distribution data in 
Fig.\ 5 of the referred paper
as a function of depletant concentration together 
with a fit $|f_{\rm ad}| \propto \gamma/\ln |\gamma|$ 
(yielding  $C_{p,m} = 0.61$ for the transition point). 
}
 \label{fig:fad}
\end{figure}

We  can also fit the experimental 
 data of Ref.\ \citenum{Welch2015} on the tail distribution 
of 
actin filaments adsorbed by a depletion attraction. 
The quantity $\zeta$ from the simple exponential 
fit (eq.\ (1) in Ref.\ \citenum{Welch2015}) 
should actually be
the inverse  correlation length $1/\xi$ and, thus, related 
to the 
free energy (per length) of adsorption by $\zeta =|f_{\rm ad}|/k_BT$
(the experimental data does not allow to determine the 
exponent $\chi$ in (\ref{ptail})). 
We obtain  free energies of adsorption as
 $|f_{\rm ad}|\simeq 1.48, 8.76, 41.38 ~ k_BT/L_p$
for the three data sets from Fig.\ 5 in   Ref.\ \citenum{Welch2015}
for depletant concentrations $C_p=0.61,0.65,0.72$, see Fig.\ \ref{fig:fad}. 
The dependence on the reduced distance to the adsorption threshold 
$\gamma \equiv (\epsilon-\epsilon_m)/\epsilon_m = (C_p-C_{p,m})/C_{p,m}$ 
is governed by the exponent $\nu$ with 
  $\nu_{SF}=1+\log$  for $\chi_{SF}=2$, see eq.\ (\ref{nu}),
i.e., $|f_{\rm ad}| \propto \gamma/\ln |\gamma|$.
The fit in Fig.\ \ref{fig:fad} shows that the data for $f_{\rm ad}$
is consistent with this scaling law.

\section[Finite polymers]{Critical potential strength  of finite polymers}
\label{sec_finite}

We have shown that tail sizes diverge with the correlation 
length $\xi$ upon approaching 
the desorption transition.  Therefore, 
we expect finite size effects as soon as the correlation length 
exceeds the polymer length, $L<\xi$. 
According to the standard argument underlying finite size effects at a
critical point, a polymer of finite length $L$ should
desorb easier, i.e., at  larger $g_c$ as soon as the 
correlation length $\xi$, which sets the scale for the desorbed tail 
length, reaches the polymer length $L$. 
In Ref.\ \citenum{Klushin2013} it has been found 
that finite flexible lattice polymers have
$g_c(L) >g_c$, i.e., finite polymers desorb easier only  for narrow 
potentials corresponding to small $\ell\ll b$, whereas 
$g_c(L) < g_c$, i.e., finite polymers adsorb easier for 
wide potentials $\ell \gg b$. The latter effect is due to the large number
of contacts with a wide attractive potential, which 
become more frequent for a  short polymer if one end is grafted to the 
potential layer. 
Therefore, the direction of the finite size shift of $g_c(L)$ is 
a subtle issue that  depends on the potential range $\ell$. 
For desorption if $\xi$ grows beyond $L$, 
finite size effects of  the  critical potential  strength $g_c$
are  calculated according to 
$\xi(g_c)=L$, which corresponds to replacing 
the free energy criterion $f_{\rm ad}(g_c)=0$ by an 
apparent offset 
$f_{\rm ad}(g_c)=k_BT/\xi(g_c)=-k_BT/L$. 
Because $\xi \propto |g-g_c|^{-\nu}$, this results in 
finite size corrections 
$g_c(L)- g_c\propto L^{-1/\nu}$, 
where the exponent $\nu$ is given by (\ref{nu}). 
This type of finite size corrections is the basis of the 
standard finite size scaling  the we employed above in eq.\ (\ref{eq:Cad})
and applies for sufficiently long polymers.

Finite size corrections have to be modified, however, 
for stiff or short  
polymers if the deflection length exceeds the polymer length, 
$L_d>L$. 
Then the polymer  acts effectively as a rigid rod, and 
internal conformational fluctuations become 
negligible, whereas global rotation or translation degrees of freedom
become relevant. These are additional fluctuation 
degrees of freedom, which tend to desorb the polymer leading to 
an {\it increase} in $g_c(L)$.  In general, 
finite size corrections from a small number of  
 global degrees of freedom can only be  of the order of
$g_c(L)- g_c\sim k_BT/L$ (eventually with logarithmic corrections).
 Therefore, this type of finite size correction is only relevant
if $\nu \le 1$, which is the case in the semiflexible and stiff limit
where $\nu_{SF}=1+{\rm log}$, see eq.\ (\ref{nu}).
For phantom flexible polymers with $\nu_F=2$ or self-avoiding chains 
with $\nu_{\rm SAW} \simeq 2.43$, on the other hand, finite size corrections 
from the diverging correlation length $\xi$ are  dominating. 
Therefore, we have to  analyze  finite size effects 
from global rotation and translation degrees of freedom 
in  the rigid and semiflexible 
limit  in the following.

\subsection{End-grafted rigid rod}

For a rigid rod, i.e., if $L_d/L\to \infty$,
the result (\ref{gcSF}) for infinite polymers 
gives $g_{c,SF} \sim k_BT/L_d \approx 0$,  i.e., 
an infinite  rigid rod will adsorb for all $g>0$
 because all internal shape fluctuations and, thus, 
entropic costs of confinement to the potential well 
are suppressed in the rigid limit. 
We neglected, however,  contributions from 
global translations and rotations  of the rigid rod
in the derivation of the adsorption threshold (\ref{gcSF}). 

For finite rigid rods, 
adsorption is not a genuine phase transition because the 
rod has only $D-1$ rotational (if the rod is axially symmetric) 
and $D$ translational 
global degrees of freedom. Nevertheless, we can 
define a characteristic potential strength for adsorption 
 either by the criterion 
that the adsorbed length $L_{\rm ad}=-{\cal H}_{\rm ad}/g$
exceeds half the polymer length\cite{Welch2015}, 
$\langle L_{\rm ad}\rangle > L/2$,
 or by a maximum in the 
second cumulant $C_{\rm ad}=
\langle {L}_{\rm ad}^2\rangle-\langle L_{\rm ad}\rangle^2$ 
 of the adsorbed length.

A  rigid rod with one  end  attached  to the boundary 
 of the attractive potential has only rotational degrees of freedom,
and  the adsorbed length can be 
calculated exactly. If $\theta$ is measured with respect to the 
positive $z$-axis, the polymer is out of the potential well 
for $0<\theta <\pi/2$ 
and inside the potential well for a small angular interval 
$0< \tilde{\theta} \equiv \pi/2-\theta<\arcsin(\ell/L)$ resulting in 
($\beta \equiv k_BT$)
\begin{align}
Z_r&=  S_D/2 + S_{D-1}\int_{0}^{\arcsin(\ell/L)} 
     d\tilde{\theta} \cos^{D-1}\tilde{\theta} e^{\beta gL}
\label{eq:Zrod}\\
  F_r &= -k_BT  \ln Z_r 
   \nonumber \\
   &\approx  -k_BT\ln \left(\frac{1}{2}S_D
      +  S_{D-1} \frac{\ell}{L} e^{\beta gL}\right)
\label{eq:Frod}
\end{align}
and
\begin{align}
 \langle L_{\rm ad}\rangle &= -\frac{\partial F_r}{\partial g} 
   =  \frac{L}{   1+
    \frac{S_D}{2 S_{D-1}}  \frac{L}{\ell} e^{-\beta gL}  }
\nonumber\\
  C_{\rm ad} &=
    -k_BT  \frac{\partial^2 F_r}{\partial g^2}  
=  L^2  \frac{ \frac{S_D}{2 S_{D-1}}  
       \frac{L}{\ell} e^{-\beta gL}}{   
   \left(1+ \frac{S_D}{2 S_{D-1}}  \frac{L}{\ell} e^{-\beta gL} \right)^2 }
\end{align}
where $S_D= 2\pi^{D/2}/\Gamma(D/2)$ is the surface 
of the unit sphere in $D$ dimensions, i.e., $S_3=4\pi$, $S_2= 2\pi$, and 
$S_1=2$ (and $\beta \equiv k_BT$). 
Both the adsorption criterion $\langle L_{\rm ad}\rangle > L/2$
and  the maximum of the second cumulant  $C_{\rm ad}$ 
agree and give 
\begin{equation}
    g_{c, {\rm rod}}(L) 
 = \frac{k_BT}{L}\ln\left(\frac{S_D}{2 S_{D-1}}  \frac{L}{\ell}\right)
\label{eq:gcrod}
\end{equation}
with $S_3/2S_2=1$ in $D=3$ and $S_2/2S_1=\pi/2$ in $D=2$.
From the derivation starting from the partition sum (\ref{eq:Zrod})
we see that $g_{c, {\rm rod}}(L)$ can also be re-written in terms of 
 the 
ratio of the accessible phase space volumes $Z_{r,0}=S_D/2$, 
where ${\cal  H}_{\rm ad}=0$ and no potential acts, and 
$Z_{r,g}= S_{D-1} \frac{\ell}{L} e^{\beta gL}$, 
where the adsorption potential acts:
\begin{equation}
   g_{c, {\rm rod}}(L) 
 = \frac{k_BT}{L}\ln\left( Z_{r,0}/Z_{r,g}(L) \right).
\label{eq:gcZ}
\end{equation}
It is important to note that  $g_{c, {\rm rod}}\approx 0$ 
for $L\to \infty$, i.e.,
desorption of a rigid rod by 
 global rotation fluctuations is  only possible for finite 
rods.
For a rigid rod,  $g_{c, {\rm rod}}(L)$   can also be 
interpreted as  the  finite size corrections
to the infinite rod result  
$g_{c,SF}  \approx 0$.

MC simulation results in Fig.\ \ref{fig:gc} 
 confirm that the adsorption threshold of 
finite polymers indeed exhibits pronounced finite size effects 
in the stiff limit and approaches  $g_{c, {\rm rod}}(L) \sim
(k_BT/L)\ln(L/\ell)$
from  eq.\ (\ref{eq:gcrod}) in the stiff limit.

\subsection{Crossover to the semiflexible regime}

Upon reducing the ratio  $L_d/L$ by 
reducing stiffness or increasing length the result 
(\ref{eq:gcrod}) for a rigid rod should cross over to the 
result (\ref{gcSF}) for an infinite  semiflexible polymer.
Equating $g_{c,{\rm rod}}(L) = g_{c,SF}\sim k_BT/L_d$ gives
$L_d\sim L$ as crossover point. For $L_d>L$ we thus expect 
adsorption of 
a  very weakly fluctuating almost rigid rod, whereas for 
$L>L_d$, there should be a crossover to the semiflexible 
result (\ref{gcSF}), for which  collisions with the  adsorption potential
 boundaries on the scale $L_d$ are important. 
This crossover proceeds via three regimes 
upon reducing the stiffness and, thus, 
the deflection length 
$L_d \sim  L_p^{1/3}\ell^{2/3}$ and the persistence length $L_p$
($L_d <L_p$ because $\ell \ll L_p$) or increasing the length $L$:

\begin{figure}[t]
\centering
 \includegraphics[width=0.4\textwidth]{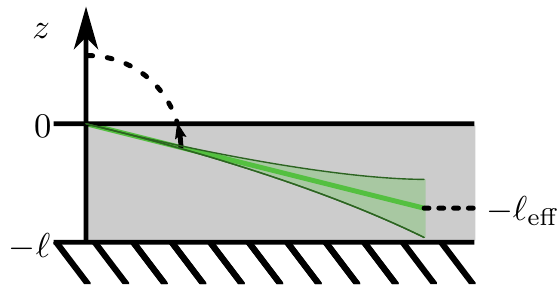}
 \caption{
 Effective thickness $\ell_{\rm eff} \approx 
 \ell - cL^{3/2}/L_p^{1/2}$ 
of a stiff rod caused by thermal fluctuations.
} 
 \label{fig:sketch}
\end{figure}

\begin{itemize}
\item[(i)] 
For large stiffness $L<L_d <L_p$  [or $N = L/b+1 < (L_p/\ell)^{1/3}(\ell/b)$] an 
adsorbed  rigid rod
starts to bend by thermal fluctuations but will have typically 
no thermal collisions with the boundaries of the 
adsorption potential of range $\ell$. Thermal fluctuations only
give rise to a finite effective 
thickness $\langle z^2\rangle^{1/2}$ of the rod. For an 
end-grafted  rod this  restricts 
the 
accessible rotation angles $\theta$ (Fig.\ \ref{fig:sketch}).

\item[(ii)] For  reduced stiffness such that $L_d<L<L_p$
 [or $(L_p/\ell)^{1/3}(\ell/b)<N< (L_p/\ell)(\ell/b)$],
 thermal fluctuations within the potential 
well are sufficient to induce 
repeated collisions with the boundaries of the adsorption layer
which gives rise to an additional  free energy cost per length.

\item[(iii)]
For $L_d<L_p<L$ [or $N>(L_p/\ell)(\ell/b)$],
the length over which the orientation of the first 
segment grafted to the potential well can persist, which is 
 by definition 
the persistence length $L_p$,  becomes smaller than the 
polymer length $L$. Then the global rotation degree of freedom 
only affects a segment of length $L_p$. 

\end{itemize}

\begin{figure*}[t]
\centering
\includegraphics[width=0.98\textwidth]{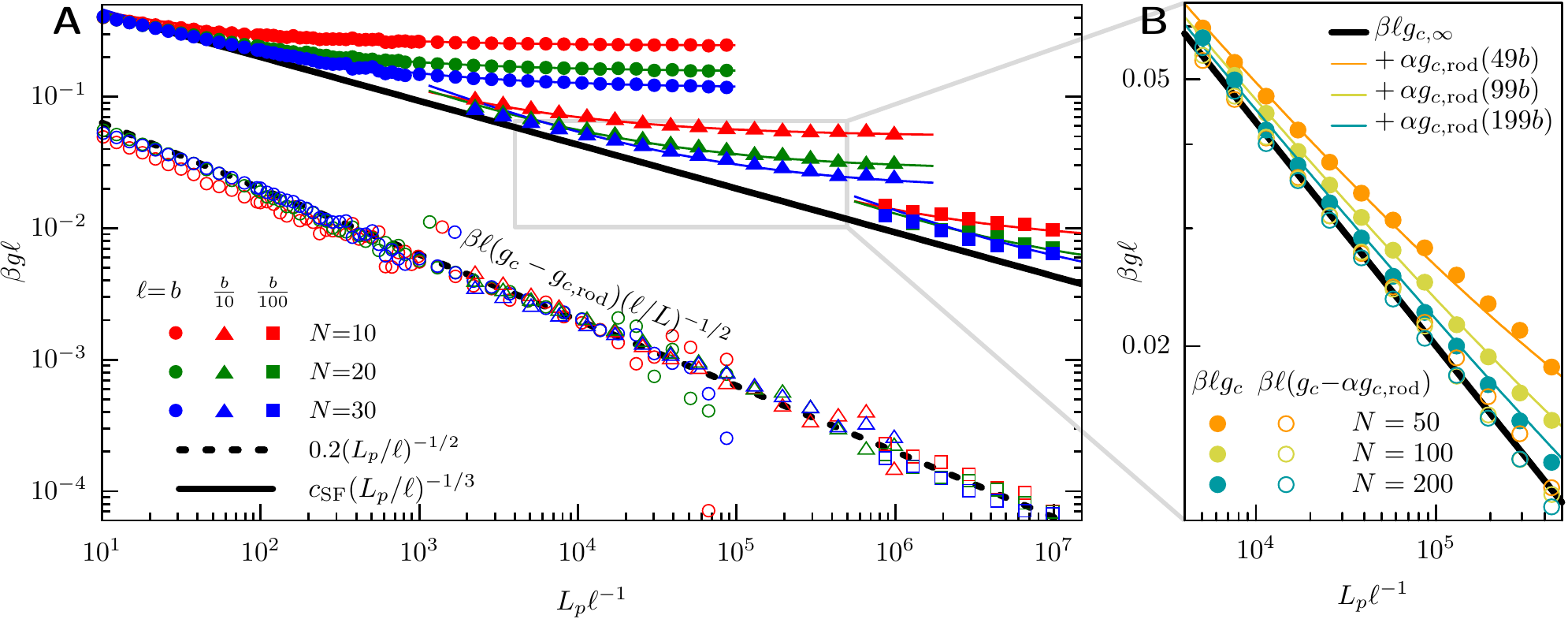}
 \caption{ 
  Finite size effects for various potential widths $\ell$ and
     contour lengths for an end-grafted semiflexible polymer. 
   For shorter chains ($N = L/b+1 = 10, 20, 30$) finite size
     effects are best described by the rigid rod with an increased effective
     thickness using eq.\ (\ref{eq:gcL1}) (\A with $\ell/b = 1,0.1,0.01$),
     which allows data collapse  onto one curve $\propto
     L_p^{{-}1/2}$.  Longer chains ($N=L/b+1= 50, 100, 200$) are best described
     by the rigid rod with additional entropic free energy cost using
     eq.\ (\ref{eq:gcrodT}) (\B with $\ell = 0.1 b$). Subtracting $\alpha
     g_{c,\rm rod}(L)$ (with $\alpha{\approx}0.5$) from the measured critical
     potential strengths recovers the critical potential strength of the
     infinite chain. The crossover takes place when $L$ becomes shorter than
     the deflection length $L_d$ (for $L<(L_p/\ell)^{1/3}(\ell/b)$).  
} 
 \label{fig:fs_long}
\end{figure*}

\begin{itemize}
\item[(i)] 
For $L<L_d$, an adsorbed  semiflexible polymer
 bends by thermal fluctuations with 
$\langle z^2\rangle \sim L^3/L_p<\ell^2$
such that it  typically  has 
no thermal collisions with the boundaries of the 
adsorption layer of size $\ell$. 
The fluctuations give the rigid rod, however, 
 an increased effective thickness
$\langle z^2\rangle^{1/2}$, which 
restricts the range of rotation angles 
where the polymer fits without bending into the potential well 
to $0< \theta-\pi/2 <\arcsin(\ell_{\rm eff}/L)$
with a reduced effective width
\begin{equation}
\ell_{\rm eff}= \ell - \langle z^2\rangle^{1/2}\approx 
 \ell - c L^{3/2}/L_p^{1/2},
\label{eq:leff}
\end{equation}
 where $c$ is 
a numerical prefactor (Fig.\ \ref{fig:sketch}).

Using the effective potential width (\ref{eq:leff}) in 
the rigid rod result  (\ref{eq:gcrod}) gives 
a $T$-dependent shift of the adsorption threshold, 
\begin{equation}
  g_{c}(L) = g_{c, {\rm rod}}(L)
+\frac{k_BT}{L} \ln\left(1+
      c \frac{L^{3/2}}{L_p^{1/2}\ell} \right).
\label{eq:gcL1}
\end{equation}
For  $L\ll L_d$ we can expand the logarithm, and 
the shift becomes
$g_c(L) - g_{c, {\rm rod}}(L) \approx c k_BT L^{1/2}/L_p^{1/2}\ell
\propto L_p^{-1/2}$.

\item[(ii)] 
If $L_d<L<L_p$, thermal fluctuations are sufficient to induce 
repeated collisions with the boundaries of the adsorption potential. 
This increases the free energy $f_{\rm ad}$ per length 
of a polymer with 
$0< \theta-\pi/2 <\arcsin(\ell/L)$ by an additional entropic 
contribution for the   confinement 
to the potential well of width $\ell$,
\begin{equation}
   f_{\rm ad} \approx -g + a k_BT \frac{1}{\ell^{2/3}L_p^{1/3}}.
\label{eq:fad}
\end{equation}
This contribution stems from restricting the internal 
shape fluctuations. 
The prefactor of the
 confinement free energy has been measured in simulations
as $a\simeq 1.1$ \cite{Bicout2001} for a hard confinement 
to a width $\ell$, which should be appropriate for large $g\gg g_c$. 
At $g=g_{c,SF}$ the free energy $f_{\rm ad}$ should vanish;
this suggests  $a=c_{SF}$ with $c_{SF}$ as in (\ref{eq:cSF})
 in the vicinity of the transition.
Both $a\simeq 1.1$ for $g\gg g_c$ and $a=c_{SF}$ for $g\approx g_c$ 
are comparable and of order unity.
The  estimate 
 (\ref{eq:fad}) is also compatible with an exponent $\nu=1$ 
of the adsorption free energy 
$|f_{\rm ad}| \sim k_BT/\xi \propto  |g-g_{c,SF}|^{\nu}$ 
close to the transition, i.e., with neglecting the logarithmic 
correction in $\nu=\nu_{SF} =  1+{\rm log}$, see eq.\ (\ref{nu}). 
Therefore, we can approximate the partition sum of the internal 
deformation degrees of freedom of an adsorbed semiflexible polymer 
by  $Z_i(L) = e^{-\beta f_{\rm ad}L}$, which replaces the rigid rod 
Boltzmann factor $e^{\beta gL}$.

Replacing $g$ in the free energy (\ref{eq:Frod}) of a rigid rod 
by  the adsorption free energy  $f_{\rm ad}$
results in a 
 critical potential strength for a finite semiflexible polymer
\begin{align}
     g_c(L)  &=  g_{c, {\rm rod}}(L)
   +  a k_BT \frac{1}{\ell^{2/3}L_p^{1/3}}\nonumber\\
   &=  g_{c, {\rm rod}}(L)+ g_{c,SF}.
 \label{eq:gcrodT}
\end{align}
This result  holds {\em independently}
of the adsorption criterion ($\langle L_{\rm ad}\rangle(g_c) = L/2$ or 
based on the second cumulant).
The 
 critical potential strength (\ref{eq:gcrodT}) can be interpreted in two ways:
first as  the rigid rod result (\ref{eq:gcrod}), which is shifted 
by an offset identical to the semiflexible result 
 $g_{c,SF}$ from eq.\ (\ref{gcSF}).
 Therefore, the temperature-induced shift with respect to the 
rigid rod result scales as 
$g_c(L) - g_{c, {\rm rod}}(L) \propto L_p^{-1/3}$,
in the regime $L<L_d$, which differs from the scaling 
$g_c(L) - g_{c, {\rm rod}}(L) \propto L_p^{-1/2}$ for $L> L_d$,
see Fig.\ \ref{fig:fs_long}.
Second as the 
 as the semiflexible result  $g_{c,SF}$  from eq.\ (\ref{gcSF}) 
for an infinite 
polymer, which  is shifted by finite size corrections due to the 
global rotation degree of freedom and  given by the 
rigid rod result $g_{c, {\rm rod}}(L)\propto L^{-1}\ln L$.

Additional finite size corrections 
arise from the internal shape fluctuations 
for $\xi>L$, which shift 
will also give a shift $\Delta g_c(L) \sim k_BT L^{-1}\ln L$ because 
$\nu =\nu_{SF}= 1+ {\rm log}$, see eq.\ (\ref{nu}).
They correspond to an additional contribution $\pm k_BT L^{-1}\ln L$ 
to $f_{\rm ad}$ in eq.\ (\ref{eq:fad}) from fluctuations in a 
finite size system. Therefore,  we expect 
$g_c(L) = \alpha g_{c, {\rm rod}}(L)+ g_{c,SF}$ with a numerical prefactor
$\alpha$ to the rigid rod correction to 
the semiflexible result  $g_{c,SF}$  from eq.\ (\ref{gcSF}).
Therefore, for  $L_p<L<L_d$,
it should be possible to find a numerical constant 
$\alpha\sim {\cal O}(1)$ such that 
$g_c(L) - \alpha g_{c, {\rm rod}}(L) \sim g_{c,SF}$ 
collapses onto the  length-independent  infinite  
semiflexible polymer result  $g_{c,SF}$  from eq.\ (\ref{gcSF}).
Our simulation data in Fig.\ \ref{fig:fs_long} is well described by 
$\alpha\simeq 0.5$ suggesting that finite size effects 
from internal fluctuations 
and from the rigid rod degrees of freedom become comparable in this 
regime.

\item[(iii)] 
If  $L_p$  is further reduced below $L$ such that 
$L>L_p$, rotations of the first segment only 
affect the polymer over a persistence length $L_p$. 
Then the free energy  is 
\begin{equation*}
   F(L) = -k_BT \ln Z(L_p) + (L-L_p) f_{\rm ad}
\end{equation*}
where $Z(L_p)$ is the partition sum of an end-grafted segment 
of length $L_p>L_d$ (as in regime (ii)) and $f_{\rm ad}$ is 
the adsorption free energy per length of the infinite polymer. 
The adsorbed length is given by 
$\langle L_{\rm ad} \rangle = - {\partial F}/{\partial g}$
and using the criterion $\langle L_{\rm ad} \rangle=L/2$ we obtain an 
adsorption threshold, which 
approaches for $L_p=L/2$ 
the semiflexible result  $g_{c,SF}$ for an infinite 
polymer.
Therefore, finite size effects from the global rotation 
degrees of freedom become irrelevant for $L_p <L/2$.   
Then we only expect small finite size effects from  internal deformation
degrees of freedom  if 
$\xi>L$, which also scale as  $\Delta g_c(L) \sim k_BT L^{-1}\ln L$ 
because  $\nu=\nu_{SF}=1+{\rm log}$.

\end{itemize}

\subsection{Semiflexible polymer confined by two hard walls}

Alternatively, we can consider confinement by a second parallel non-adsorbing
hard wall at a distance $L_z >\ell$ in order to avoid that the polymer
diffuses away.  This type of confinement has been considered in
Ref.\ \citenum{Welch2015}.  For a completely rigid rod we calculate the ratio
of the phase space volumes $Z_{r,0}$, where ${\cal H}_{\rm ad}=0$, and
$Z_{r,g}$, where the adsorption potential acts, and then apply
eq.\ (\ref{eq:gcZ}) as above for the end-grafted rod. Contrary to the
  end-grafted polymer, where the first bead is held fixed, the prefactor for a
  free polymer is $\propto 1/ Nb$, since the potential acts on $N=(L+1)/b$
  beads, where the entropic confinement depends on the actual contour length
  $L = (N-1)b$.

\begin{figure}[t]
\centering
 \includegraphics[width=0.46\textwidth]{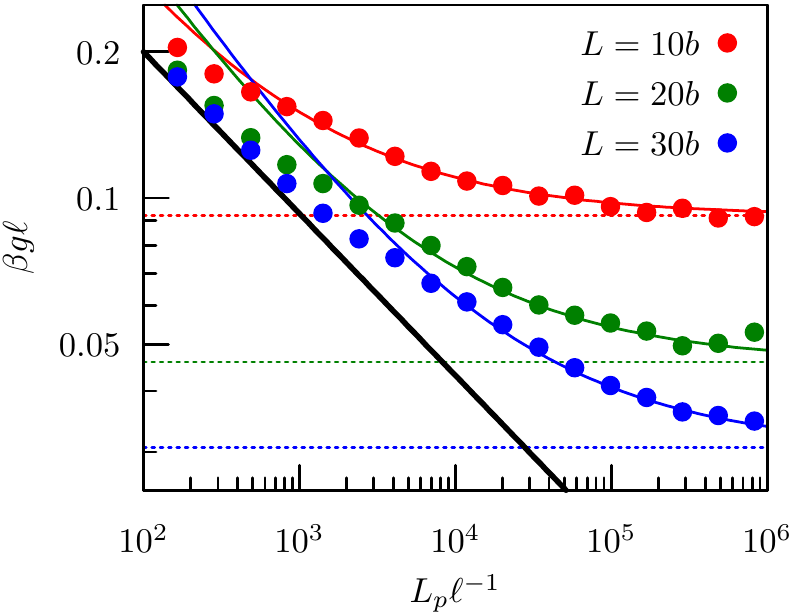}
 \caption{ \eqref{eq:gcrodwall} 
  Finite size effects for a free
     semiflexible polymer (with $N=L/b+1 = 10, 20, 30$) confined between two
     walls separated by a distance $L_z = 10b$. In front of one wall there is
     an attractive square-well potential with width $\ell = 0.1b$. The
     critical potential strength is best described by eq.\ \eqref{eq:gcL1wall}
     (solid curves), which shows the crossover from an infinite 
  semiflexible  chain
     (eq.\ \eqref{gcSF}, solid black line) to a rigid rod
     (eq.\ \eqref{eq:gcrodwall}, dashed horizontal lines).  
}
 \label{fig:gcwall}
\end{figure}

Now the rod can perform global rotations {\it and} translations. 
We parametrize global rotations by the  angle $\theta$  
(or $\tilde{\theta} \equiv \pi/2-\theta$ 
with $0<\theta,\tilde{\theta}<\pi/2$ for non-polar rods) 
and global translations by the coordinate $z$ of the rod center
(rotations within the adsorbing plane or translations parallel to the 
adsorbing plane play no role).
The phase space volume $Z_{r,0}$ is (for $\ell \ll L_z$) simply the 
partition sum of a free rod between two walls with distance $L_z$,
which is obtained from observing that only angles 
$0< \tilde{\theta} <\arcsin(L_z/L)$ are possible and 
that, for a given angle $\tilde{\theta}$, only $z$-coordinates 
$z>(\sin\tilde{\theta}) L/2$ and $z<L_z - (\sin\tilde{\theta}) L/2$ 
are accessible,
\begin{align}
 Z_{r,0} &=  S_{D-1}\int_{0}^{\arcsin{L_z/L}}
     d\tilde{\theta} \cos^{D-1}\tilde{\theta}
      (L_z - L \sin\tilde{\theta}) \nonumber \\
   &\approx S_{D-1} \frac{L_z^2}{2L}.
\label{eq:Zrodwall}
\end{align}
Likewise, phase space volume  $Z_{r,g}$ is approximately the partition 
sum of a polymer confined between two walls with separation $\ell$, 
i.e., $Z_{r,g} \approx  S_{D-1} {\ell^2}/{2L}$
resulting in 
\begin{equation}
   g_{c, {\rm rod}}(L) 
 = \frac{k_BT}{Nb}\ln\left( Z_{r,0}/Z_{r,g}(L) \right)
\approx  \frac{2k_BT}{Nb}\ln\left(  \frac{L_z}{\ell} \right)
\label{eq:gcrodwall}
\end{equation}
(which deviates by a factor of 2 from the result given in 
 Ref.\ \citenum{Welch2015}).
 MC simulation results in Fig.\ \ref{fig:gcwall} 
 confirm that the adsorption threshold of 
finite polymers between two hard walls 
 indeed approaches  $g_{c, {\rm rod}}(L)$
from  eq.\ (\ref{eq:gcrodwall}) in the stiff limit.

For $L_z\sim L$ the result becomes similar to (\ref{eq:gcrod}) for 
 an end-grafted polymer. The global degrees of freedom involved in
(\ref{eq:gcrodwall}) are, however,  rotations {\it and} translations, whereas 
they are only rotations for the end-grafted rod. 
Also corrections for finite temperatures are similar 
and go through the same three regimes for decreasing 
$L_p$.

\begin{itemize}

\item[(i)] For  $L<L_d$ the polymer is a weakly fluctuating 
rigid rod with an 
 increased effective thickness leading to an 
effectively decreased potential width
$\ell_{\rm eff}$ as in eq.\ (\ref{eq:leff}.
Using  $\ell_{\rm eff}$   in 
the rigid rod result  (\ref{eq:gcrodwall}) gives a 
$T$-dependent shift
\begin{equation}
  g_{c}(L) = g_{c, {\rm rod}}(L)
+\frac{2k_BT}{L} \ln\left(1+
          c \frac{L^{3/2}}{L_p^{1/2}\ell} \right),
\label{eq:gcL1wall}
\end{equation}
which, for $L\ll L_d$ gives a shift 
$g_c(L) - g_{c, {\rm rod}}(L) \sim    k_BT L^{1/2}/L_p^{1/2}\ell
\propto L_p^{-1/2}$ as 
 as for end-grafting.

\item[(ii)] For  $L_d<L<L_p$, there is 
an additional entropic free energy cost due to repeated
collisions with the boundaries, see eq.\ (\ref{eq:fad}). 
This leads to the same shift as in the end-grafting 
result (\ref{eq:gcrodT}).

\item[(iii)] for $L>L_p$ the  overall orientation 
of the polymer is lost and finite 
size effects only come from the global translational 
degree of freedom. 
If the size of the polymer as measured by its 
end-to-end distance $\langle R^2 \rangle^{1/2}$ is of the 
order of $L_z$ or larger, we do not expect finite size corrections
from global translation. 
For $L_z>\langle R^2 \rangle^{1/2}$,
 finite size corrections from global translation 
will be of the order of 
$g_c(L) - g_{c,SF} \sim k_BT \ln(L_z/\langle R^2\rangle^{1/2})$.
\end{itemize}

\subsection{Finite size effects in the flexible limit}

As discussed above, in the flexible regime $L_p\ll \ell$
 finite size 
corrections from internal shape fluctuations result 
in a shift $|g_c(L)- g_c|\propto L^{-1/\nu}$ with $\nu=\nu_F=2$ 
if $\xi>L$. 

There is, however,  an additional source
of finite size corrections associated with a finite 
potential range $\ell$.  
Upon decreasing $L_p$, the mean square radius 
$\langle R^2 \rangle \sim  2L_p L$ as given by the 
flexible chain result with an effective segment length of $b\approx 2L_p$ 
becomes smaller than the 
the square of the potential range for $L < \ell^2/2L_p$. 
Then, the entire chain can accommodate into 
the potential well without entropic free energy costs
resulting in $g_c \approx 0$ for $L < \ell^2/2L_p$.
Therefore, in the flexible limit, the finite size result 
 for $g_c$  should approach the infinite
polymer result $g_{c,F}$ from eq.\ (\ref{gcF}) from below.

\section{Finite size scaling procedure}

Based on our  results for the finite size corrections 
of the adsorption threshold, we obtain a method to analyze 
adsorption data for finite semiflexible polymers
such as filamentous actin.
We assume that the adsorption threshold $g_c(L)$ has been determined
experimentally or in simulations for finite polymers 
and want to demonstrate
how to fit to the theory presented above, which will allow 
us to extract possible fit parameters such as the persistence length $L_p$ 
or the potential range $\ell$. 

Our above results 
in  the semiflexible regime  $L<L_p$
(eqs.\ (\ref{eq:gcrod}), (\ref{eq:gcL1}) for case (i) $L<L_d$   and 
eqs.\ (\ref{eq:gcrodwall}), 
(\ref{eq:gcL1wall}) for case (ii) $L_d<L<L_p$) show that 
global rotation and translational degrees of freedom play a 
dominating role for finite size corrections in this regime. 
In order to correct for these effects we can subtract the 
rigid rod result $g_{c, {\rm rod}}(L)$ and continue with 
an analysis of  the data for $g_c(L) - g_{c, {\rm rod}}(L)$. 

We then have to distinguish 
between case  (i) $L<L_d$  and case (ii) $L_d<L<L_p$.
In case (i) we fit the shifted data $g_c(L) - g_{c, {\rm rod}}(L)$
according to (\ref{eq:gcrod}) and  (\ref{eq:gcrodwall}) 
with $g_c(L) - g_{c, {\rm rod}}(L)\sim k_BT L^{1/2}/L_p^{1/2}\ell$
for both end-grafting and wall-confinement. 
In case (ii) we fit the shifted data $g_c(L) - g_{c, {\rm rod}}(L)$
according to (\ref{eq:gcL1}) and 
(\ref{eq:gcL1wall}) using 
$g_c(L) - g_{c, {\rm rod}}(L)\sim  g_{c,SF}\sim k_BT /{\ell^{2/3}L_p^{1/3}}$.
In both cases, these fits should  enable us to extract material 
parameters such as the persistence length $L_p$.

A similar fit procedure (using only case (ii))
has actually been used in Ref.\ \citenum{Welch2015}
to analyze data but on phenomenological  grounds. 
The arguments presented in this paper systematically 
justify this technique and show the necessary distinction between 
case (i) of an essentially rigid rod for $L>L_d$ and case (ii) 
of a semiflexible polymer for $L_d<L<L_p$. 
In  Ref.\ \citenum{Welch2015} the use of case (ii) was appropriate 
 because the potential range $\ell$
and, thus, $L_d$ was  small.

\section{Discussion and Conclusion}

In this paper we unraveled the different 
adsorption regimes for finite semiflexible polymers 
if persistence length $L_p$, potential range $\ell$, and 
the finite contour length $L$ are changed. 
An overview of all regimes is given in table \ref{tab:regimes}.
Finite semiflexible polymers exhibit  three distinct  regimes 
for the adsorption potential strength:
(i) a  flexible or 
Gaussian regime if the persistence length is smaller than the adsorption 
potential range, (ii) a semiflexible regime if the persistence length is 
larger than the potential range, and (iii) for finite polymers, 
a novel crossover to a  rigid rod 
regime if the  deflection length exceeds the contour length.

Our main result is the novel adsorption regime (iii)
for finite stiff polymers if the deflection length $L_d$ 
exceeds the 
contour length,  $L<L_d\sim L_p^{1/3}\ell^{2/3}$, see Fig.\ \ref{fig:gc}.
Then the adsorption threshold is governed by the global 
rotational and translational degrees of freedom of a 
finite rigid rod. 
For end-grafted polymers 
we find in the rigid rod limit 
$g_{c, {\rm rod}}(L) \sim (k_BT/L)\ln(L/\ell)$,
see eq.\ (\ref{eq:gcrod}).
Upon reducing the stiffness or increasing the length,
the threshold crosses over to the 
semiflexible regime with $g_{c,SF}  \sim k_BT/L_d \sim
k_BT/L_p^{1/3}\ell^{2/3}$
according to (\ref{gcSF}), which can be described by 
eq.\ (\ref{eq:gcrodT}).
For adsorption in confinement between two walls we find 
analogous result, see Fig.\ \ref{fig:gcwall}.
Based on our results we can derive a finite size scaling 
procedure to analyze adsorption data on finite semiflexible 
polymers.

In Ref.\ \citenum{Welch2015} the adsorption of 
the semiflexible polymer F-actin  has been studied recently. 
For F-actin contour and persistence lengths 
$L\sim L_p\sim 10-20 {\rm\mu m}$ are typical. 
Depletion potentials in Ref.\ \citenum{Welch2015} 
have a range $\ell\sim 10 {\rm nm}$. 
Other possible attractive potentials are 
electrostatic interactions with
$\ell\sim 1 {\rm nm}$ at physiological conditions for 
monovalent ions and 
larger ranges $\ell \propto 1/z\sqrt{c_{\rm salt}}$ at 
lower salt concentrations or higher valencies $z$, 
which gives similar ranges as for depletion. 
Therefore, we are in the regime 
$L_d \ll L_p \sim L$ for F-actin experiments and 
the adsorption threshold should be described by 
the semiflexible result for an infinite 
polymer, $g_{c,SF}  \sim k_BT/L_d \sim
k_BT/L_p^{1/3}\ell^{2/3}$
according to (\ref{gcSF}). 
Only relatively small finite size  
corrections of the rigid rod form according to (\ref{eq:gcrodT}) 
should be observable according to our theory.
This is in accordance with the results in Ref.\ \citenum{Welch2015}.

The novel rigid rod regime
with a pronounced length-dependence of the  adsorption 
threshold (\ref{eq:gcrod}) should be accessible, for example,  for 
short microtubules.
 For  a microtubule persistence length 
 $L_p  \sim 5 {\rm mm}$ and 
similar potential ranges $\ell \sim 10{\rm nm}$ as for F-actin, 
we find  $L_d \sim L \ll L_p$ for contour lengths 
$L\sim 1{\rm \mu m}$.

Moreover, we presented a theory for the loop and tail distributions
of flexible and semiflexible polymers 
and the critical exponents governing these distributions 
close to the adsorption threshold. 
Our results (\ref{ploop}) for loops and (\ref{ptail}) for tails
 explain that, close to the transition, 
 semiflexible polymers have significantly smaller loops and 
both flexible and semiflexible polymers desorb by expanding their 
tail length. 
This agrees with simulation observations in  
Refs.\ \citenum{Hsu2013,Welch2015}.
The  tail distribution 
allows us to directly extract the free energy per length 
 of adsorption $f_{\rm ad}$ from the experimental data 
presented in Ref.\ \citenum{Welch2015}, see Fig.\ \ref{fig:fad}.

\end{document}